\documentclass[twocolumn,twocolappendix]{aastex631}

\usepackage{color}
\usepackage{amsmath}
\usepackage{multirow}
\usepackage{booktabs}
\usepackage[utf8]{inputenc}
\usepackage{appendix}

\usepackage{subfigure}
\usepackage{graphicx}

\usepackage{float}

\newcommand{\angstrom}{\textup{\AA}}
\newcommand{\myand}{\;\, \& \;\,}

\newcommand{\LCDM}{$\Lambda$CDM\,}

\newcommand{\Ob}{\Omega_b}
\newcommand{\Om}{\Omega_m}
\newcommand{\Ol}{\Omega_{\Lambda}}

\newcommand{\sig}{\sigma_8}

\begin{document}

\title{Testing Lyman Alpha Emitters and Lyman-Break Galaxies as Tracers of Large-Scale Structures at High Redshifts}

\correspondingauthor{Ho Seong Hwang}
\email{hhwang@astro.snu.ac.kr}

\author[0009-0003-9748-4194]{Sang Hyeok Im}
\affiliation{Department of Physics and Astronomy, Seoul National University, 1 Gwanak-ro, Gwanak-gu, Seoul 08826, Republic of Korea}

\author[0000-0003-3428-7612]{Ho Seong Hwang}
\affiliation{Department of Physics and Astronomy, Seoul National University, 1 Gwanak-ro, Gwanak-gu, Seoul 08826, Republic of Korea}
\affiliation{SNU Astronomy Research Center, Seoul National University, 1 Gwanak-ro, Gwanak-gu, Seoul 08826, Republic of Korea}

\author[0000-0003-3095-6137]{Jaehong Park}
\affiliation{Korea Institute for Advanced Study, 85 Hoegi-ro, Dongdaemun-gu, Seoul 02455, Republic of Korea}
\affiliation{Space Science Exploration Directorate, Korea AeroSpace Administration (KASA), 537, Haeansaneop-ro, Sanam-myeon, Sacheon-si, Gyeongsangnam-do, Republic of Korea}

\author[0000-0002-6810-1778]{Jaehyun Lee}
\affiliation{Korea Astronomy and Space Science Institute, 776 Daedeokdae-ro, Yuseong-gu, Daejeon 34055, Republic of Korea}

\author[0000-0002-4362-4070]{Hyunmi Song}
\affiliation{Department of Astronomy and Space Science, Chungnam National University, 99 Daehak-ro, Yuseong-gu, Daejeon, 34134, Republic of Korea}

\author[0000-0001-8227-9516]{Stephen Appleby}
\affiliation{Asia Pacific Center for Theoretical Physics, Pohang, 37673, Republic of Korea}
\affiliation{Department of Physics, POSTECH, Pohang 37673, Republic of Korea}

\author[0000-0003-0225-6387]{Yohan Dubois}
\affiliation{CNRS and Sorbonne Université, UMR 7095, Institut d'Astrophysique de Paris, 98 bis, Boulevard Arago, F-75014 Paris, France}

\author[0000-0003-3940-7687]{C. Gareth Few}
\affiliation{E.A. Milne Centre for Astrophysics, University of Hull, Hull HU6 7RX, UK}

\author[0000-0003-4446-3130]{Brad K. Gibson}
\affiliation{Woodmansey Primary School, Hull Road, Woodmansey, HU17 0TH, UK}

\author[0000-0002-4391-2275]{Juhan Kim}
\affiliation{Center for Advanced Computation, Korea Institute for Advanced Study, 85 Hoegiro, Dongdaemun-gu, Seoul 02455, Republic of Korea}

\author[0000-0003-4164-5414]{Yonghwi Kim}
\affiliation{Korea Institute of Science and Technology Information, 245 Daehak-ro, Yuseong-gu, Daejeon, 34141, Republic of Korea}

\author[0000-0001-9521-6397]{Changbom Park}
\affiliation{Korea Institute for Advanced Study, 85 Hoegi-ro, Dongdaemun-gu, Seoul 02455, Republic of Korea}

\author[0000-0003-0695-6735]{Christophe Pichon}
\affiliation{Korea Institute for Advanced Study, 85 Hoegi-ro, Dongdaemun-gu, Seoul 02455, Republic of Korea}
\affiliation{CNRS and Sorbonne Université, UMR 7095, Institut d’Astrophysique de Paris, 98 bis, Boulevard Arago, F-75014 Paris, France}
\affiliation{IPhT, DRF-INP, UMR 3680, CEA, L’Orme des Merisiers, Bât 774, F-91191 Gif-sur-Yvette, France}

\author[0000-0001-5135-1693]{Jihye Shin}
\affiliation{Korea Astronomy and Space Science Institute, 776 Daedeokdae-ro, Yuseong-gu, Daejeon 34055, Republic of Korea}

\author{Owain N. Snaith}
\affiliation{Department of Physics and Astronomy, University of Exeter, Exeter EX4 4QL, UK}

\author[0000-0003-0570-785X]{Maria Celeste Artale}
\affiliation{Departamento de Ciencias Fisicas, Universidad Andres Bello, Fernandez Concha 700, Las Condes, Santiago, Chile}

\author[0000-0003-1530-8713]{Eric Gawiser}
\affiliation{Department of Physics and Astronomy, Rutgers, the State University of New Jersey, Piscataway, NJ 08854, USA}

\author[0000-0002-4902-0075]{Lucia Guaita}
\affiliation{Universidad Andres Bello, Facultad de Ciencias Exactas, Departamento de Fisica, Instituto de Astrofisica, Fernandez Concha 700, Las Condes, Santiago RM, Chile}

\author[0000-0002-2770-808X]{Woong-Seob Jeong}
\affiliation{Korea Astronomy and Space Science Institute, 776 Daedeokdae-ro, Yuseong-gu, Daejeon 34055, Republic of Korea}

\author[0000-0003-3004-9596]{Kyoung-Soo Lee}
\affiliation{Department of Physics and Astronomy, Purdue University, 525 Northwestern Ave., West Lafayette, IN 47906, USA}

\author[0000-0001-9850-9419]{Nelson Padilla}
\affiliation{Instituto de Astronomía Teórica y Experimental (IATE), CONICET-UNC, Laprida 854, X500BGR, Córdoba, Argentina}

\author[0000-0002-9176-7252]{Vandana Ramakrishnan}
\affiliation{Department of Physics and Astronomy, Purdue University, 525 Northwestern Avenue, West Lafayette, IN 47906, USA}

\author[0000-0001-6162-3023]{Paulina Troncoso}
\affiliation{Escuela de Ingeniería, Universidad Central de Chile, Avenida Francisco de Aguirre 0405, 171-0614 La Serena, Coquimbo, Chile}

\author[0000-0003-3078-2763]{Yujin Yang}
\affiliation{Korea Astronomy and Space Science Institute, 776 Daedeokdae-ro, Yuseong-gu, Daejeon 34055, Republic of Korea}

\begin{abstract}
We test whether Lyman alpha emitters (LAEs) and Lyman-break galaxies (LBGs) can be good tracers of high-z large-scale structures, using the Horizon Run 5 cosmological hydrodynamical simulation. We identify LAEs using the Ly$\alpha$ emission line luminosity and its equivalent width, and LBGs using the broad-band magnitudes at $z\sim2.4$, $3.1$, and $4.5$. We first compare the spatial distributions of LAEs, LBGs, all galaxies, and dark matter around the filamentary structures defined by dark matter. The comparison shows that both LAEs and LBGs are more concentrated toward the dark matter filaments than dark matter. We also find an empirical fitting formula for the vertical density profile of filaments as a binomial power-law relation of the distance to the filaments. We then compare the spatial distributions of the samples around the filaments defined by themselves. LAEs and LBGs are again more concentrated toward their filaments than dark matter. We also find the overall consistency between filamentary structures defined by LAEs, LBGs, and dark matter, with the median spatial offsets that are smaller than the mean separation of the sample. These results support the idea that the LAEs and LBGs could be good tracers of large-scale structures of dark matter at high redshifts.
\end{abstract}

\keywords{Large-scale structure of the universe, Hydrodynamical simulations, Lyman-alpha galaxies, Lyman-break galaxies}

\section{Introduction} \label{sec:intro}

Studying the physical properties of large-scale structures in the universe is important for constraining cosmological models (e.g., \citealt{Park_1990, Lin_1996, Hawkins2003, Eisenstein_2005, Percival_2007, Park_2012ApJL,  Hong_2020, Dong_2023}). Since the first large redshift surveys of galaxies in the 1970s and 1980s (e.g., \citealt{Gregory_Thompson_1978, Kirshner_1978, Davis_1982, Geller_Huchra_1989}), the spatial distribution of galaxies has been a useful tool for studying of cosmology and the structure formation. Statistical analyses of the galaxy distribution (e.g., two-point and angular correlation functions; \citealt{Peebles_1975,Landy_Szalay_1993,Hamilton_1993}) allow us to make a direct comparison between observations and theoretical predictions from cosmological simulations. However, most of the early observational studies of galaxy distribution were conducted with single slit spectroscopy and were therefore inefficient in terms of the number of galaxies observed simultaneously. 

Thanks to the development of the multi-object spectrographs in the 1990s, which allow simultaneous observations for hundreds of galaxies, the number of galaxies with measured redshifts has increased rapidly (e.g., \citealt{Shectman_1996, York_2000, Colless_2001, Hwang_2016}). In addition, the appearance of advanced detectors and larger telescopes enables deeper redshift surveys over larger areas. Therefore, we are now in a good position to easily measure the spatial distribution of high-redshift galaxies, which is important for understanding cosmology and structure formation (e.g., \citealt{Lilly_2009, Silverman_2015, Sohn_2021}).

Despite these improvements in the observing techniques, spectroscopic observations for a large number of galaxies still require a large amount of telescope time and effort. The situation becomes worse if one tries to observe galaxies only in a certain redshift range, because one cannot know the exact redshift of a galaxy before obtaining the spectrum of the galaxy. In this regard, narrow-band imaging observations for Lyman alpha emitters \citep[LAEs;][]{Hu_McMahont1996, Ouchi_2003, Gawiser_2007, Lee_2014} and the drop-out technique for Lyman-break galaxies \citep[LBGs;][]{Steidel_1998, Giavalisco_2004, Bouwens_2007, Toshikawa_2016} are common ways to select galaxies within within a certain redshift range at high redshifts without spectroscopic observations. One can reduce the number of target galaxies for follow-up spectroscopic observations using these kinds of approaches, making the redshift survey of high redshift galaxies more efficient. Especially the typical redshift range of LAEs from narrow-band imaging observations ($\Delta z \lesssim 0.04-0.08$), which is much smaller than that of LBGs from using the drop-out technique ($\Delta z \sim 0.5$), enables us to select galaxies within a narrow redshift range very efficiently. 

One of the ongoing observational projects using such narrow-band imaging to study galaxy distribution at high redshifts is the One-hundred-deg$^2$ DECam Imaging in Narrow-bands (ODIN) survey (\citealt{Lee_2024}; \citealt{Ramakrishnan_2023}). We will use the data from this survey to identify LAEs at three redshifts $z\sim2.4$, $3.1$, and $4.5$ with three custom narrow-band filters that have central wavelengths of $419$, $501$, and $673\;\rm{nm}$. We will also use these LAEs to identify and study the large-scale structure\footnote{The distribution of galaxies or dark matter larger than galaxy cluster scale ($\gtrsim$ a few $\rm{Mpc}$).}. Because the target redshifts of the ODIN survey are around the epoch of peak mass accretion in galaxy clusters, the data will play an important role in improving our understanding of cosmic structure formation and evolution. The observation is currently ongoing with the Dark Energy Camera (DECam) mounted at Víctor M. Blanco 4-m telescope of Cerro Tololo Inter-American Observatory. The very first paper about the relation between Ly$\alpha$ blobs and the distribution of LAEs from the ODIN observed data of the COSMOS field at $z\sim3.1$ has recently been published \citep{Ramakrishnan_2023}.

Although using LAEs and LBGs is a convenient method to identify high-z galaxies, they are a subset of the entire galaxy populations. Therefore, it may not be obvious whether they can represent well the spatial distribution of the entire galaxy population or dark matter at high redshifts. In this study, we would like to examine the capability of LAEs and LBGs as tracers of the large-scale structures at high redshifts, by considering the large-scale structures of dark matter as a reference. We focus on the filamentary structures of LAEs, LBGs, all galaxies, and dark matter using the data from a new cosmological hydrodynamical simulation, the Horizon Run 5 \citep{Lee2021, Park_2022}. We will examine the spatial distribution of these tracers around filaments and how those tracers give different shapes of filamentary structures. Please note that our analyses are in three-dimensional space, which may not be sufficient for direct comparison with observations. We plan to present a direct comparison with observations using two-dimensional mock data as a future study.

In Section \ref{sec:data}, we describe the Horizon Run 5 data along with the selection methods for LAE and LBG samples. Section \ref{sec:results} includes our results on the filamentary structures of our samples and their spatial distribution around the filaments. In Section \ref{sec:discussion}, we compare our results with previous studies and discuss the capability of LAEs and LBGs as tracers of large-scale structures of dark matter at high redshifts. Finally, we present the summary in Section \ref{sec:conclusion}.

\section{Data and Sample Selection}\label{sec:data}

\subsection{Data}\label{subsec:data}

We use the data of the Horizon Run 5 \citep[HR5:][]{Lee2021} cosmological hydrodynamical simulation for our analysis. HR5 has a simulation box size\footnote{Throughout this paper, we present all spatial quantities in comoving scale.} of $\sim 1\;\rm{cGpc}^3$ with a high-resolution cuboid zoom-in region of $1,049 \times 119 \times 127 \; (\rm{cMpc})^3$, where the spatial resolution reaches down to $\sim1\;\rm{proper \; kpc}$. This unique geometry of the zoom-in region is designed to be optimized for generating mock lightcone data of deep field surveys such as ODIN \citep{Lee_2024}, LSST \citep{LSST2012}, and DESI \citep{DESI_2016}. The linear power spectrum and the initial conditions are generated at $z=200$ and evolved using the adaptive mesh refinement code RAMSES \citep{Teyssier_2002} until $z=0.625$. Various subgrid physics are also implemented during the simulation to model galaxy formation and evolution. After the simulation, galaxies (i.g., subhalos) are identified using PGalF, which is an extended galaxy finder based on the friends-of-friends algorithm (see Appendix A of \citealt{Kim_2023} for details). The physical properties of each galaxy are calculated with the particles belonging to the galaxy. For more details about the HR5 simulation, please see \citealt{Lee2021}. We use three snapshot data of HR5 at $z\sim2.4$, $3.1$, and $4.5$, which are similar to the three target redshifts of the ODIN survey. Throughout HR5 simulation and this study, we adopt the \LCDM cosmology with $\Om = 0.3$, $\Ol = 0.7$, $\Ob = 0.047$, $\sig = 0.816$, and $h_0 = 0.684$ as compatible with Planck data \citep{Planck_2016}. 

HR5 has particles and grids with different resolutions to simultaneously achieve the high spatial resolution at the zoom-in region and obtain the structures with very large scales ($\sim 1\;\rm{cGpc}$) throughout the simulation box. Those particles and grids with different resolutions may be mixed near the boundary between the regions with the different resolutions. This can affect the estimated physical properties of the halos and subhalos (i.e., galaxies) near the boundary. To deal with this issue, we use only the galaxies and dark matter particles in the $90\times90\times1000$ $(\rm{cMpc})^3$ region at the center of the highest resolution region. Hereafter, we will refer to this region as the central zoom-in region. Inside this region, all galaxies and dark matter particles are at least $3\;\rm{cMpc}$ away from the lower resolution particles.

We generate the samples of dark matter particles and all galaxies at the central zoom-in region for three redshifts: $z\sim2.4$, $3.1$, and $4.5$. We randomly select $0.1\%$ of the total dark matter particles to reduce the computational load of further analysis (e.g., generating a density map or extracting filaments from those particles). We also apply a minimum stellar mass cut of $2.137 \times 10^8 \; \rm{M_{\odot}}$, corresponding to $\sim100$ star particles, to the galaxy samples to select galaxies with statistically reliable physical properties. We use the same stellar mass cut when generating the LAE and LBG samples in Section \ref{subsec:LAB_LBG_selection}. The sample of all galaxies above the mass threshold is divided further into several subsamples with different stellar mass ranges. The stellar mass range of each subsample is available in Table \ref{table:num_samples}. These ranges are selected to make the number of galaxies in each mass bin at least larger than $10,000$ and to be similar to each mass bin in each redshift. From these subsamples, we examine how the stellar masses of galaxies affect their spatial distribution and resulting filamentary structures. In Table \ref{table:num_samples}, we present the number of dark matter particles and galaxies in each sample and subsample.

\subsection{LAE and LBG Selection}\label{subsec:LAB_LBG_selection}

\subsubsection{LAEs}\label{subsubsec:LAE_selection}

We assign the Ly$\alpha$ emission line luminosity ($\rm{L}_{Ly\alpha}$) and the rest-frame equivalent width ($\rm{REW}$) of each galaxy to select LAEs (\citealt{Weinberger_2019}; Park et al., in preparation). To achieve this, we use the conditional probability function of $\rm{REW}$ given the UV luminosity, motivated by the empirical models (e.g., \citealt{Dijkstra_Wyithe_2012}; \citealt{Gronke_2015}; \citealt{Oyarzun_2017}; \citealt{Mason_2018}). In this study, we adopt the equation (4) of \cite{Mason_2018}, and adjust the free parameters to match the observed Ly$\alpha$ luminosity functions at $z\sim2.4$, $3.1$, and $4.5$ (Park et al., in preparation). We then assign $\rm{REW}$ value to each galaxy according to the UV luminosity of the galaxy and the conditional probability that we have calculated. The $\rm{REW}$ value of each galaxy is then converted to $\rm{L}_{Ly\alpha}$ using the following equation:
\begin{equation}
\rm{L}_{Ly\alpha} = \rm{REW} \times \rm{L}_{\rm{UV}} \times \frac{\nu_{\alpha}}{\lambda_{\alpha}}{\left( \frac{\lambda_{\alpha}}{\lambda_{UV}}\right)}^{-\beta -2},
\end{equation}
where, $\nu_{\alpha}$ and $\lambda_{\alpha}$ are the frequency and wavelength of the Ly$\alpha$ emission line, respectively. The $\rm{L}_{UV}$ represents the UV luminosity, and $\lambda_{UV}$ is the wavelength of the UV emission. Here, we assume that the UV continuum of each galaxy follows a power law with a slope of $\beta$. This semi-empirical modeling does not capture the exact physical conditions of each galaxy as is done with radiative transfer calculations (e.g., \citealt{Zheng_2014}; \citealt{Behrens_2018}). However, it allows us to populate LAEs in a large simulation box and to study their statistical properties such as luminosity function and spatial distribution.

We summarize the selection criteria for LAEs at $z\sim2.4$, $3.1$, and $4.5$ in equations (\ref{eq:LAE_selection_2.4}), (\ref{eq:LAE_selection_3.1}), and (\ref{eq:LAE_selection_4.5}), respectively. Here, we consider the $5\sigma$ detection limit (\citealt{Lee_2024}) of the three narrow-band filters of the ODIN survey as the minimum $\rm{L}_{Ly\alpha}$ values. We also give a minimum REW value of $20\;\angstrom$ to select those galaxies with significant emission line luminosity compared to the continuum level. We adopt these minimum REW values from the LAE selection criteria of the ODIN survey (see \citealt{Ramakrishnan_2023} and \citealt{Firestone_2023arXiv}). 
\begin{equation}\label{eq:LAE_selection_2.4}
z \sim 2.4:\;
\rm{log}\left(\frac{L_{Ly\alpha}}{erg \; s^{-1}}\right) > 42.18 \myand \rm{REW} > 20 \;\angstrom,
\end{equation}

\begin{equation}\label{eq:LAE_selection_3.1}
z \sim 3.1: \; \rm{log}\left(\frac{L_{Ly\alpha}}{erg \; s^{-1}}\right) > 42.20 \myand \rm{REW} > 20  \;\angstrom,
\end{equation}

\begin{equation}\label{eq:LAE_selection_4.5}
z \sim 4.5: \; \rm{log}\left(\frac{L_{Ly\alpha}}{erg \; s^{-1}}\right) > 42.39 \myand \rm{REW} > 20  \;\angstrom.
\end{equation}
Finally, $\sim3.8$, $7.4$, and $8.1 \%$ of galaxies with stellar masses larger than $2.137 \times 10^8\;\rm{M_{\odot}}$ are selected as LAEs for $z\sim2.4$, $3.1$, and $4.5$, respectively. We mainly use the sample of all LAEs at each redshift, but divide it into three stellar mass bins if necessary. The relevant sample sizes and the mass ranges of the LAE samples and their subsamples are summarized in Table \ref{table:num_samples}.

\subsubsection{LBGs}\label{subsubsec:LBG_selection}

We select LBGs using the selection criteria of apparent magnitudes of galaxies as done for observations \citep[e.g., ][]{Steidel_2003, Toshikawa_2016, Kawamata_2018}. In this study, we are interested in comparing LBGs, LAEs, all galaxies, and dark matter particles at each redshift. Thus, we use the snapshot data of HR5 for each redshift instead of lightcone data with continuous redshift ranges. Therefore, it is important to note that LBG samples in our study cannot be directly compared to photometrically selected LBGs from observations, which have redshift ranges of $\Delta z \sim 0.5$.

\begin{figure}[t!]
    \centering
    \includegraphics[width=\linewidth]{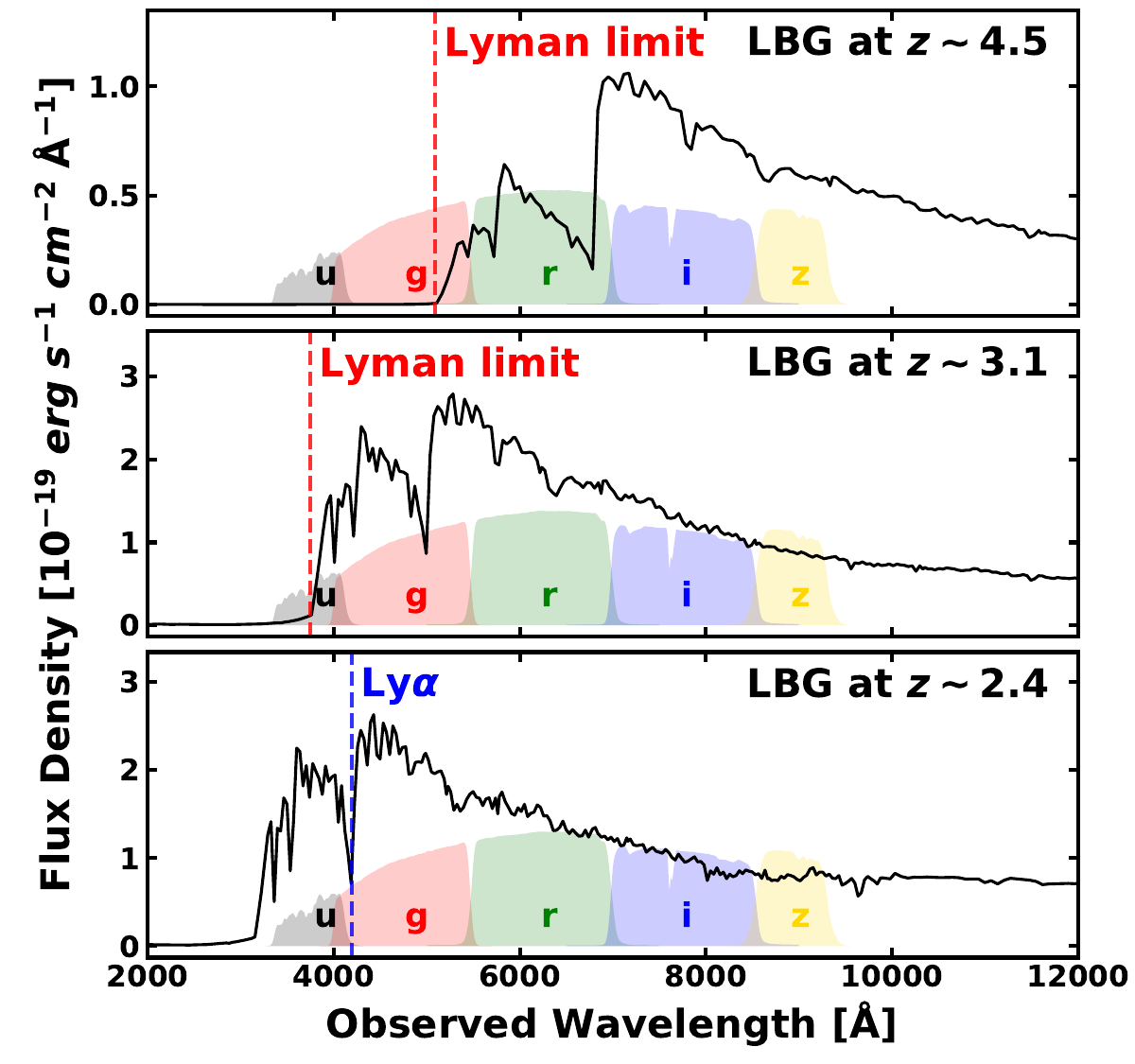}
    \caption{Example spectral energy distributions of LBG at z $\sim 2.4$, $3.1$, and $4.5$ which are selected using the criteria of equations (\ref{eq:LBG_selection_2.4}), (\ref{eq:LBG_selection_3.1}), and (\ref{eq:LBG_selection_4.5}), respectively. They show clear Lyman break features for $z\sim3.1$ and $4.5$, and the suppressed $u$-band fluxes due to the Ly$\alpha$ forest (see text for details). For comparison, we also present the filter response functions for the broad-band filters of the Subaru Strategic Program in arbitrary units.}
    \label{fig:LBG_SEDs}
\end{figure}

\begin{deluxetable*}{c|l|rrr|c|l|rrr}
    \tabletypesize{\scriptsize}
    \renewcommand{\arraystretch}{1.5}
    \renewcommand{\tabcolsep}{1.5mm}
    \tablecaption{The number of galaxies or dark matter particles in each sample for three redshifts. The stellar mass ranges for the subsamples of galaxy samples are also represented. Note that LAEs and LBGs are also included in all galaxy samples. We also present the number of galaxies which are selected as both LAE and LBG. \label{table:num_samples}}
    \tablehead{\colhead{} & \colhead{} & \colhead{$z\sim4.5$} & \colhead{$z\sim3.1$} & \colhead{$z\sim2.4$} & \colhead{} & \colhead{} & \colhead{$z\sim4.5$} & \colhead{$z\sim3.1$} & \colhead{$z\sim2.4$}}
    \startdata
    \multicolumn{2}{l|}{DM particles ($0.1\%$ sampled)} & 456,716 & 453,766 & 450,738 & \multirow{4}{*}{LAEs} & $2.137 \times 10^8 < \rm{M_{\bigstar}/M_{\odot}} < 1.5 \times {10}^9$ & 3,904 & 6,245 & 1,590 \\ \cline{1-5}
    \multirow{5}{*}{All Galaxies} & $2.137 \times 10^8 < \rm{M_{\bigstar}/M_{\odot}} < 3.5 \times {10}^8$ & 29,289 & 70,113 & 72,651 & & $1.5 \times 10^9 < \rm{M_{\bigstar}/M_{\odot}} < 2.5 \times {10}^9$ & 1,379 & 3,850 & 1,620 \\
    & $3.5 \times {10}^8 < \rm{M_{\bigstar}/M_{\odot}} < 6.0 \times {10}^8 $ & 23,233 & 63,961 & 76,826 & & $2.5 \times 10^9 < \rm{M_{\bigstar}/M_{\odot}}$ & 1,833 & 9,457 & 9,505 \\
    & $6.0 \times {10}^8 < \rm{M_{\bigstar}/M_{\odot}} < 1.5 \times {10}^9 $ & 22,700 & 72,912 & 97,243 & & \multicolumn{1}{c|}{total} & 7,116 & 19,552 & 12,715 \\ \cline{6-10}
    & $1.5 \times {10}^9 < \rm{M_{\bigstar}/M_{\odot}}$ & 12,335 & 56,784 & 87,905 & \multirow{4}{*}{LBGs} & $2.137 \times 10^8 < \rm{M_{\bigstar}/M_{\odot}} < 1.5 \times {10}^9$ & 2,486 & 5,693 & 1,292 \\ 
    & \multicolumn{1}{c|}{total} & 87,557 & 263,770 & 334,625 & & $1.5 \times 10^9 < \rm{M_{\bigstar}/M_{\odot}} < 2.5 \times {10}^9$ & 3,374 & 6,961 & 3,402 \\
    & \; & \; & \; & \; & \; & $1.5 \times {10}^9 < \rm{M_{\bigstar}/M_{\odot}}$ & 5,664 & 17,996 & 12,411 \\
    & \; & \; & \; & \; & \; & \multicolumn{1}{c|}{total} & 11,524 & 30,650 & 17,105 \\ \cline{6-10}
    & \; & \; & \; & \; & \multicolumn{2}{c|}{LAEs $\rm{\cap}$ LBGs} & 3,179 & 8,679 & 2,740 \\ 
    \hline
    \enddata
\end{deluxetable*}

To calculate the apparent magnitudes of each galaxy, we first model the spectral energy distribution (SED) of the galaxy. To do that, we consider each star particle belonging to a galaxy as a simple stellar population (SSP). Then we calculate the intrinsic SED of each galaxy as the summation of the SEDs of all the star particles and apply the Milky Way dust attenuation model to the intrinsic SED. The SSP model and the dust attenuation law are selected to give the well-matched prediction of observational properties such as UV luminosity function and dust mass - stellar mass relation. The details about the choice of models for SSP and dust attenuation law will be provided by Song et al., (in preparation). We apply the \cite{madau1995} IGM absorption model to the intrinsic SED of each galaxy to obtain the observed SED. The apparent magnitudes of each galaxy are calculated by integrating the SED over the filter set of Subaru Strategic Program \citep[SSP:][]{Aihara2018a, Aihara2018b, Aihara_2022} that we are going to use for the analysis of the ODIN data (\citealt{Ramakrishnan_2023}; \citealt{Firestone_2023arXiv}). Here, we calculate all the magnitudes in the AB magnitude system \citep{Oke1983}.

\begin{figure*}[t!]
    \centering
    \includegraphics[width=\linewidth]{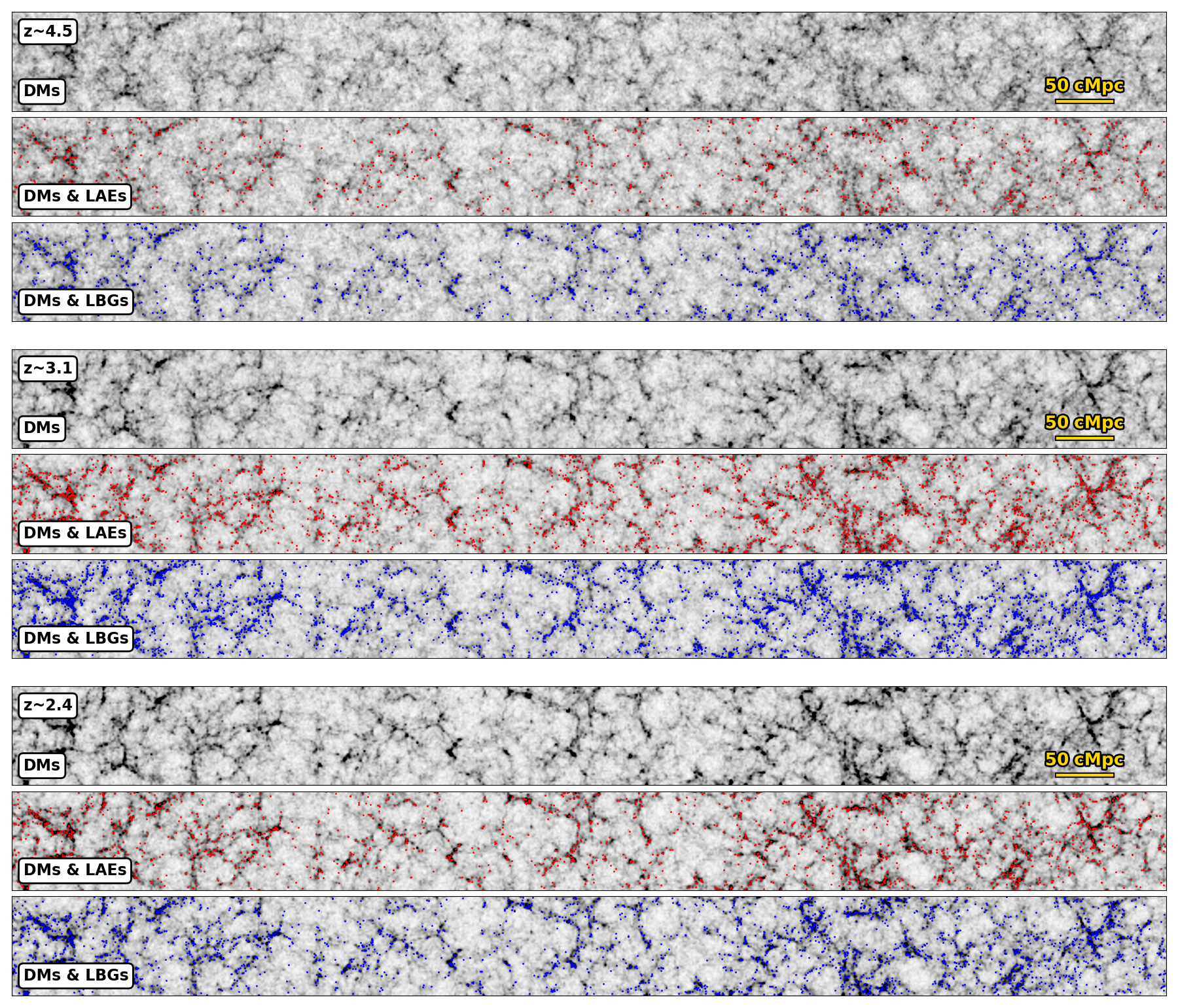}
    \caption{Spatial distribution of dark matter particles, LAEs and LBGs at $z\sim4.5$ (top three panels), $3.1$ (middle three panels), and $2.4$ (bottom three panels). For visualization, only a slice with the line-of-sight thickness of $15 \; \rm{cMpc}$ is shown. The yellow horizontal bar at the upper panel of each redshift represents the comoving length of $50\;\rm{Mpc}$. LAEs and LBGs seem to follow well the dense regions of dark matter.}
    \label{fig:2d_distribution_of_samples}
\end{figure*}

We present the selection criteria for LBGs at $z\sim2.4$, $3.1$, and $4.5$ in equations (\ref{eq:LBG_selection_2.4}), (\ref{eq:LBG_selection_3.1}), and (\ref{eq:LBG_selection_4.5}), respectively. The basic idea of the $z\sim3.1$ and $4.5$ criteria is to identify $u$- and $g$-dropout galaxies using the magnitude difference across the Lyman limit. However, the Lyman limit for $z\sim2.4$ is outside the $u$-band, so our criteria for $z\sim2.4$ is set to capture the suppressed $u$-band magnitude due to the Ly$\alpha$ forest absorption lines. Therefore, we set the upper limit of the $u-g$ color for $z\sim2.4$. These selection criteria are optimized to make the selection efficiency peaks at each ODIN target redshift. For three redshifts, we also consider the $5\sigma$ limiting magnitudes of SSP as the maximum broad-band magnitudes. These values are measured with randomly placed $2''$ diameter apertures on the SSP image of the COSMOS field (see Section 2.1 of \citealt{Ramakrishnan_2023} for details). With these criteria, $\sim5.11$, $11.6$, and $13.2\%$ of galaxies with stellar masses larger than $2.137 \times 10^8\;\rm{M_{\odot}}$ are selected as LBGs for $z\sim2.4$, $3.1$, and $4.5$, respectively. The LBG sample is further divided into three subsamples based on their stellar masses. Similar to the LAEs, we mainly use the sample of all LBGs at each redshift unless it is mentioned. The total number of selected LBGs and the mass ranges of the subsamples for three redshifts are summarized in Table \ref{table:num_samples}. We also present the example SEDs of our LBGs for each redshift in Figure \ref{fig:LBG_SEDs}. They show clear Lyman break features (for $z\sim3.1$ and $4.5$) or suppressed fluxes in the $u$-band (for $z\sim2.4$). It should be noted that some galaxies in our simulation could be selected as both LAEs and LBGs; we also list the number of such galaxies in Table \ref{table:num_samples}.

\begin{align} 
& z \sim 2.4 : \label{eq:LBG_selection_2.4} \\ 
&0.6<(u - g)<1.2 \myand -0.2 < (r - i) < 0.1 \myand \nonumber \\
& (u - r) > 2.1\times(r - i) + 0.5 \myand (g - r) < 0.7 \myand \nonumber \\
&(u - r) < 2.1\times(r - i) + 1.1 \myand \nonumber \\
& g < 26.3 \myand r < 26.0 \myand i < 25.9, \nonumber \\
\nonumber \\ 
& z \sim 3.1 : \label{eq:LBG_selection_3.1} \\ 
& (u - g) > 1.2 \myand (r - i) < 0.3 \myand \nonumber \\
& 0.1 < (g - r) < 0.7 \myand (u - g) > 2.5\times(g - r) + 0.5 \myand  \nonumber \\ 
& g < 26.3 \myand r < 26.0 \myand i < 25.9, \nonumber \\ \nonumber
\\
& z \sim 4.5 : \label{eq:LBG_selection_4.5} \\ 
& (g - i) > 2.3 \myand (g - i) > 2.1 \times (i - z) + 2.2 \myand \nonumber \\
& (r - i) > 0.8 \myand (g - i) > 0.4 \times (i - z) + 1.2 \myand \nonumber \\
& -0.5 < (i - z) < 0.3 \myand i < 25.9. \nonumber
\end{align}

\begin{figure*}[t!]
    \centering
    \includegraphics[width=\textwidth]{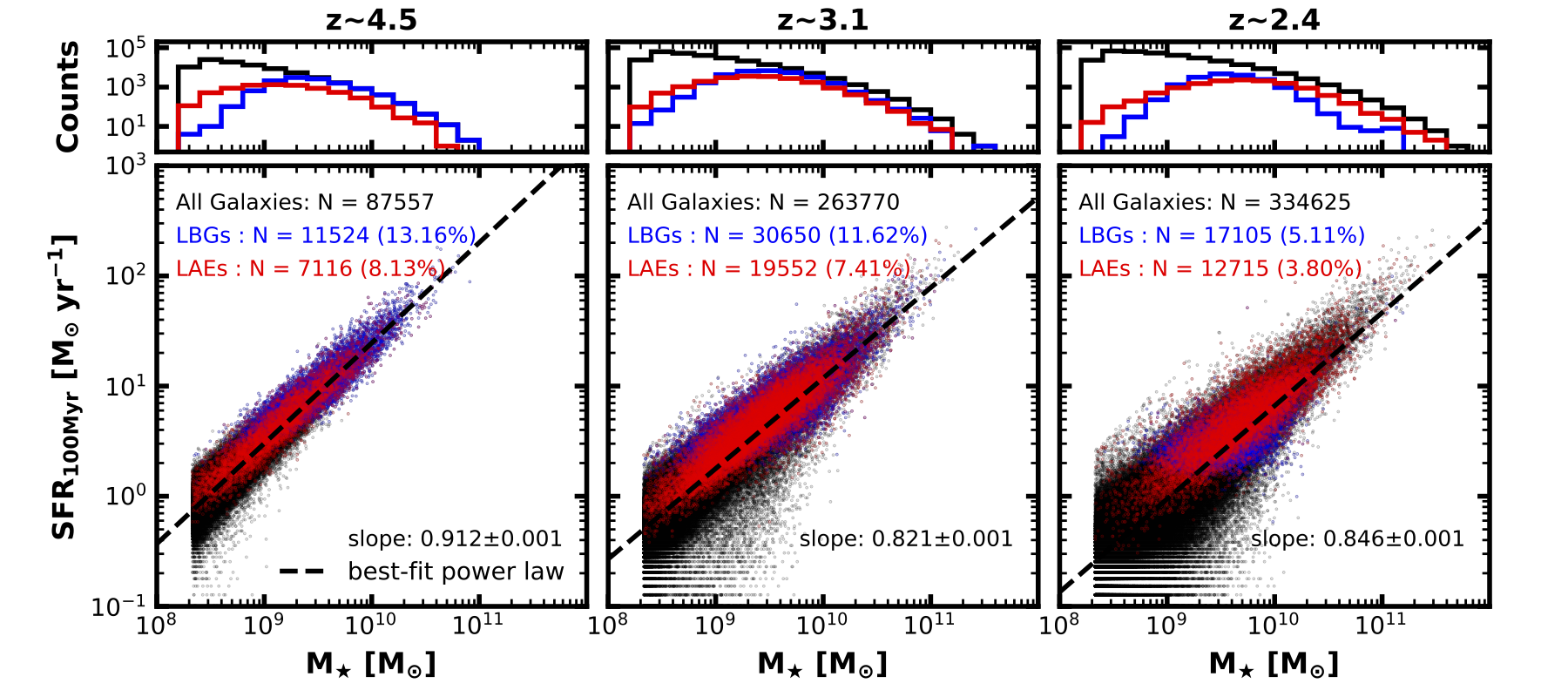}
    \caption{(upper panels) Stellar mass distributions of LAE, LBG, and all galaxy samples. (lower panels) The relations between stellar mass and star formation rate of the samples. Here, we use the average star formation rate over $100\; \rm{Myr}$ to refer to the star formation rate of each galaxy. The best-fit power law slopes for all galaxy samples are also presented.}
    \label{fig:SFR_vs_Mstar}
\end{figure*}

\begin{figure*}[t!]
    \centering
    \includegraphics[width=\textwidth]{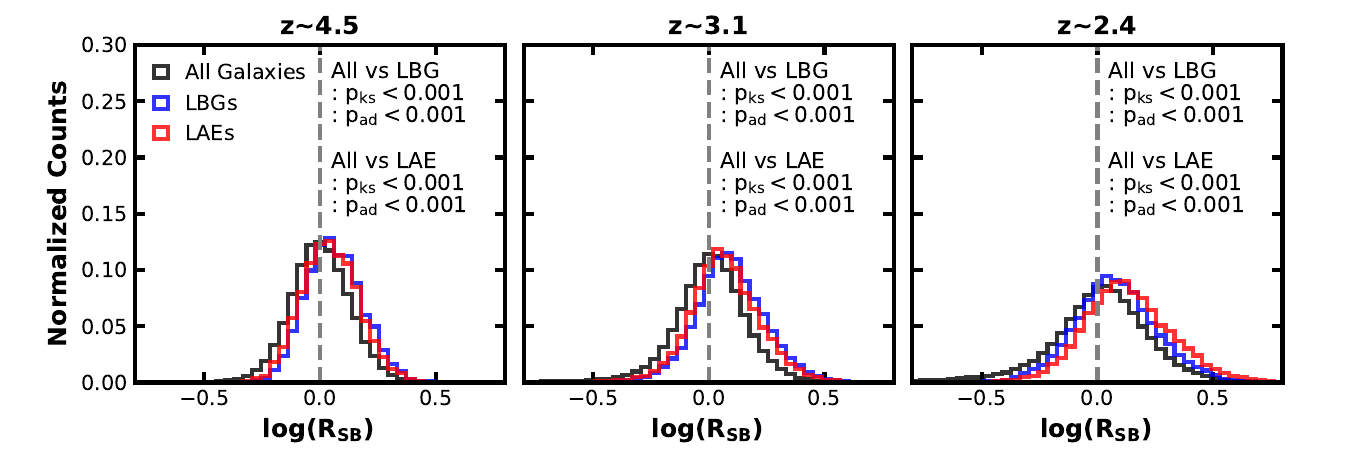}
    \caption{Distributions of starburstiness ($R_{\rm{SB}}$) of the LAE, LBG, and all galaxy samples at three redshifts. LAE and LBG samples show more skewed distributions toward larger $R_{\rm{SB}}$ value than the entire galaxy population at three redshifts. The p-values of the KS test and AD test between LAE, LBG, and all galaxy samples at three redshifts are also presented in each panel.}
    \label{fig:Rsb_comparison}
\end{figure*}


\subsection{Physical Properties of the Sample Galaxies}

\subsubsection{Spatial distribution}\label{subsec:spatial_distribution}

We examine the spatial distribution of our samples of LAEs, LBGs, and dark matter particles. For convenience, we plot the two-dimensional distribution for our samples in a sliced region with the line-of-sight thickness of $15 \; \rm{cMpc}$ of the central zoom-in region at $ z\sim2.4$, $3.1$, and $4.5$ in Figure \ref{fig:2d_distribution_of_samples}. We present the distribution of dark matter particles at the background of each panel to compare it with the distribution of LAEs and LBGs. The plot clearly shows that LAEs and LBGs trace the dense regions of dark matter particles at the three redshifts, as expected. We can also see the growth of structures as universe evolves.

\subsubsection{Relation between star formation rate \rm{\&} stellar mass}\label{subsec:SFR_vs_Mstar}

To have an idea about the difference in physical properties between our samples, we examine the relation between star formation rate (SFR) and stellar mass of LAE, LBG, and all galaxy samples. In Figure \ref{fig:SFR_vs_Mstar}, we present the relation for the three redshifts along with the stellar mass distribution of our samples. Here, we use the average star formation rate over $100\; \rm{Myr}$ ($\rm{SFR}_{100\;\rm{Myr}}$) to refer to the star formation rate of each galaxy. The relations between SFR and stellar mass of all galaxies are well fitted with a single power law at three redshifts as well known in previous studies (e.g., \citealt{Elbaz_2007}; \citealt{Speagle_2014}; \citealt{Salmon_2015}; \citealt{Iyer_2018}). Here, we perform least-squares fitting and obtain the best-fit power law slopes of $0.85$, $0.82$, and $0.91$ for $z\sim2.4$, $ 3.1$, and $4.5$, respectively.

\begin{figure*}
    \centering
    \includegraphics[width=\linewidth]{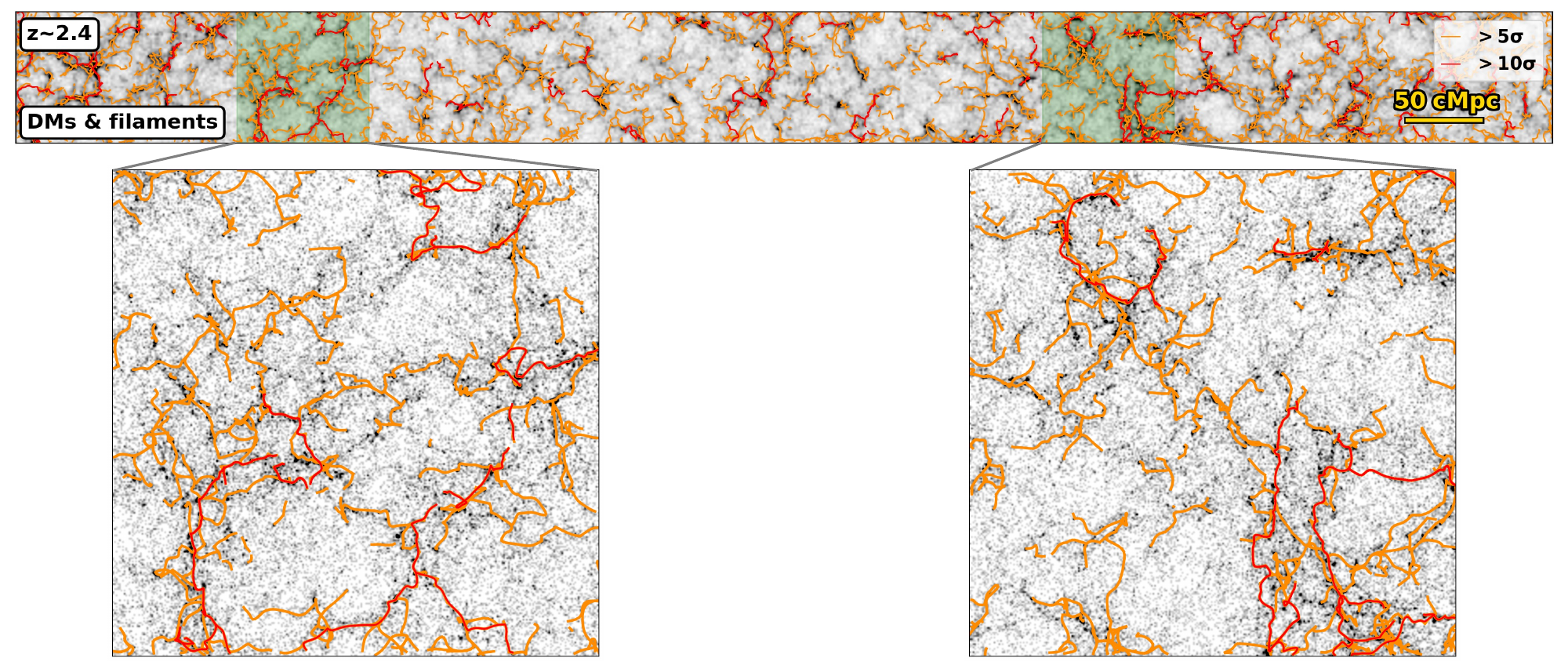}
    \caption{Filamentary structures defined by dark matter particles at $z\sim2.4$ with the minimum persistence level of $5\sigma$ (orange) and $10\sigma$ (red).  For convenience, only a slice with the line-of-sight thickness of $15\;\rm{cMpc}$ is presented. We note that we use three-dimensional filamentary structures for our analysis, while only the projected view of them is shown in here. Please also note that filaments with a minimum persistence level of $5\sigma$ include those with $10\sigma$.}
    \label{fig:DM_filaments}
\end{figure*}

To quantitatively compare the star formation rates among the samples without the dependence of stellar mass on the SFR, we calculate  the starburstiness \citep[$R_{\rm{SB}}$;][]{Elbaz_2011} of each galaxy using the following equation:
\begin{equation}
    R_{\rm{SB}} = \frac{\rm{sSFR}}{\rm{sSFR}_{MS}}.
\end{equation}
Here, $\rm{sSFR}$ is the specific star formation rate of each galaxy, and ${\rm{sSFR}}_{\rm{MS}}$ is the $\rm{sSFR}$ of the star-forming main-sequence galaxy with the same stellar mass. We use the best-fit power law relation between $\rm{SFR}$ and stellar mass from the least-squares fitting, to calculate the ${\rm{sSFR}}_{\rm{MS}}$ of each galaxy. We then compare the distribution of $R_{\rm{SB}}$ of the LAE, LBG, and all galaxy samples in Figure \ref{fig:Rsb_comparison}. LAE and LBG samples have more skewed distributions of $R_{\rm{SB}}$ toward larger values than all galaxy samples. We also perform the Kolmogorov-Smirnov (KS) test and the Anderson-Darling (AD) k-sample tests on the $R_{\rm{SB}}$ distributions. The results are shown in each panel of Figure \ref{fig:Rsb_comparison} with the p-values indicating the probability that the two samples are drawn from the same parent distribution. All the p-values are small (i.e., $<0.001$), indicating that the LAEs and LBGs are distinctive from the entire galaxy samples. This result, along with the skewed $R_{\rm{SB}}$ distributions of LAEs and LBGs in Figure \ref{fig:Rsb_comparison}, implies that LAEs and LBGs tend to be more star-forming than the entire galaxy populations at the three redshifts.

\begin{figure*}
    \centering
    \includegraphics[width=\textwidth]{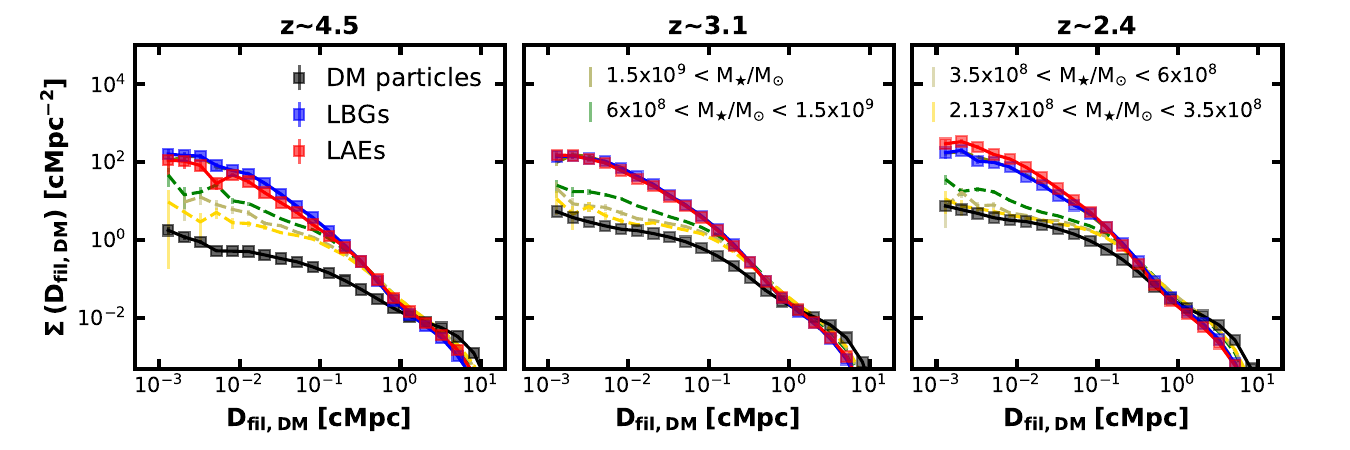}
    \caption{Vertical density profile around dark matter filaments as a function of distance to dark matter filaments ($D_{fil,\rm{DM}}$) for each sample at three redshifts. Dashed lines in each panel represent the subsamples of the entire galaxies with different stellar mass ranges (see labels in the middle and right panel). The errorbars represent the standard deviations from bootstrapping with 1,000 times. For three redshifts, LAEs, LBGs, and all galaxies are more concentrated toward the dark matter filaments than dark matter particles. Galaxies with larger stellar masses are also more concentrated than those with lower stellar masses.}
    \label{fig:Dskel_distribution}
\end{figure*}

\section{Results}\label{sec:results}

\subsection{Filamentary Structures}\label{subsec:extracting_filaments}

To identify the filamentary structures of our samples, we apply the Discrete Persistent Structure Extractor code\footnote{\tt http://www2.iap.fr/users/sousbie/web/html/indexd41d.html} \citep[DisPerSE;][]{Sousbie2011a, Sousbie2011b} to those samples. A brief description of how DisPerSE extracts filamentary structure is as follows. First, DisPerSE generates a density map of the sample using the Delaunay tessellation. Then, DisPerSE finds sets of small segments, connecting saddle points and local maxima of the density field. These sets of segments are extracted as the filamentary structure of the sample. More details on DisPerSE are available in \cite{Sousbie2011a} and \cite{Sousbie2011b}.

When running DisPerSE, the lack of data points outside the boundary would affect the density field estimation and the resulting filamentary structures near the boundary. DisPerSE provides several options to solve this problem by creating artificial data points outside the boundary. We test these options of DisPerSE and find that each option gives slightly different filamentary structures near the boundary. To be conservative, we decide not to use the filamentary structures within $5\; \rm{cMpc}$ from the boundary for our analysis.

Another important parameter to consider is the choice of the minimum persistence level of the extracted filamentary structures. The persistence level of a filament is defined as the density difference between its end points at local maxima and saddle point. The filaments with larger persistence levels are considered more robust features than those with smaller persistence levels. We can extract more robust but less detailed filamentary structures by selecting those with higher persistence levels. There are many studies based on DisPerSE, which adopt $3\sigma$ of the noise level of random data as the minimum persistence level (e.g., \citealt{Malavasi2016, Kraljic2017}). This means that the selected filamentary structures would not be generated by random fluctuations within $3\sigma$ confidence level. However, we use a higher level ($5\sigma$) for dark matter particles because $3\sigma$ gives much more detailed filamentary structures than the distribution of the galaxy samples (i.e., LAEs, LBGs, and all galaxies within stellar mass ranges). Figure \ref{fig:DM_filaments} shows the extracted filamentary structures of dark matter particles at $z\sim2.4$. For comparison, we present the filamentary structures with different persistence levels as different colors (see labels in Figure \ref{fig:DM_filaments}).

\subsection{Spatial Distributions of Samples Around Filaments defined by Dark Matter}\label{subsec:Dskel_dist_DM_filament}

We compare the spatial distributions of LAEs, LBGs, all galaxies, and dark matter particles around the filamentary structures defined by dark matter particles. We first calculate the distance from a galaxy or a dark matter particle to the closest filament for each sample. We will refer to this as $D_{fil,\rm{DM}}$; the distance to the filamentary structures defined by dark matter. We then calculate the vertical density profile around the filamentary structures using the following equation:
\begin{flalign}\label{eq:pdf_calculation_formula}
    \Sigma(D_{fil,\rm{DM}}) = & \; \frac{{\rm \#\,of\,objects\,at\,}r_i < D_{fil,\rm{DM}} < r_j}{\pi({r_j}^2 - {r_i}^2)} \; \times \frac{1}{N_{\rm{tot}}}.
\end{flalign}
Here, $\Sigma \, (D_{fil,\rm{DM}})$ is the two-dimensional probability density that a galaxy (or a particle) is located at the annulus between $r_i$ and $r_j$ from the filament, and $N_{tot}$ is the total number of data in the sample. 

\begin{figure*}[t!]
    \centering
    \includegraphics[width=\textwidth]{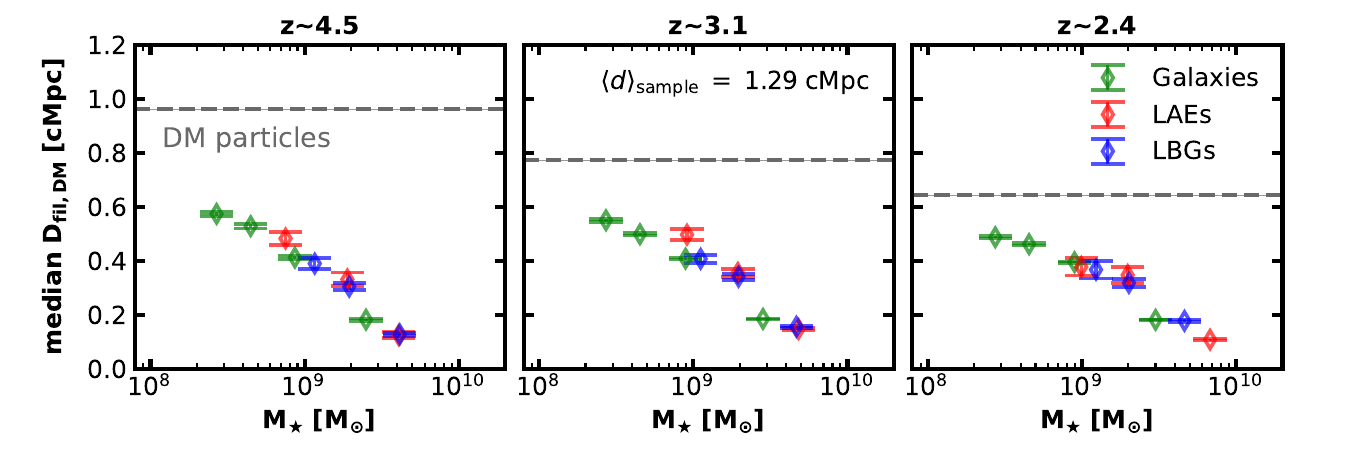}
    \caption{Median $D_{fil,\rm{DM}}$ (i.e., distance to dark matter filament) of each sample at three redshifts. Vertical errorbars and shaded regions around dashed lines represent $1\sigma$ errors from bootstrapping with 1,000 times. The LAE, LBG, and galaxy samples show smaller median values than dark matter particles. In the right panel, we also present the mean separation of the sample used for extracting filaments.}
    \label{fig:median_Dskel_comparison}
\end{figure*}

In this calculation, the galaxies and dark matter particles close to the nodes (i.e., the intersections of the filaments) can affect the calculated density profiles around the filamentary structures. To remove such effects, we exclude the galaxies and dark matter particles close ($< 2\;\rm{cMpc}$) to the nodes when calculating $\Sigma \, (D_{fil,\rm{DM}})$. We choose the minimum separation of $2\;\rm{cMpc}$ as the distance from filaments where the density profiles of our samples become smaller than those of randomly distributed particles (see Section \ref{subsec:fitting_formula}).

Figure \ref{fig:Dskel_distribution} shows the $\Sigma \, (D_{fil,\rm{DM}})$ of each sample at three redshifts. LAEs, LBGs, and all galaxies show steeper slopes and are more concentrated toward the filamentary structure defined by dark matter particles than dark matter particles themselves. This result is consistent with the visual impression of Figure \ref{fig:2d_distribution_of_samples}, where LAEs, LBGs, and all galaxies reside in the dense regions of dark matter particles. In addition, the galaxies with larger stellar masses are more concentrated toward the filaments than those with smaller stellar masses, indicating that more massive galaxies would form and evolve in the denser region of dark matter. These trends appear at all three redshifts.

Next, we compare the median values of $D_{fil,\rm{DM}}$ of our samples in Figure \ref{fig:median_Dskel_comparison}. Here, we use only the galaxies and dark matter particles with  $D_{fil,\rm{DM}}<2\;\rm{cMpc}$ to consider those directly relevant to filamentary structures as described in Section \ref{subsec:fitting_formula}. Here, we use the mass-binned subsamples of LAEs and LBGs at each redshift. LAEs, LBGs, and all galaxies have smaller median values than dark matter particles, regardless of redshift. This result again shows that LAEs, LBGs, and all galaxies are more concentrated toward the filamentary structures defined by dark matter than dark matter particles themselves. In addition, more massive galaxies have smaller median values than less massive galaxies, indicating that galaxies with larger stellar masses are more concentrated toward dark matter filaments than those with smaller stellar masses. It is also important to note that LAEs and LBGs follow well the trend of the entire galaxy population (i.e., decreasing median value as stellar mass increases). However, the absolute values of median $D_{fil,\rm{DM}}$ of the samples would differ for different choices of the minimum persistence level or the size of the sample size used to extract the filaments. Therefore, we also present the mean separation, ${\langle d \, \rangle}_{\rm{sample}}$, of the tracer sample (here, dark matter particles) for comparison with other studies. 

\begin{figure*}[t!]
    \centering
    \includegraphics[width=\textwidth]{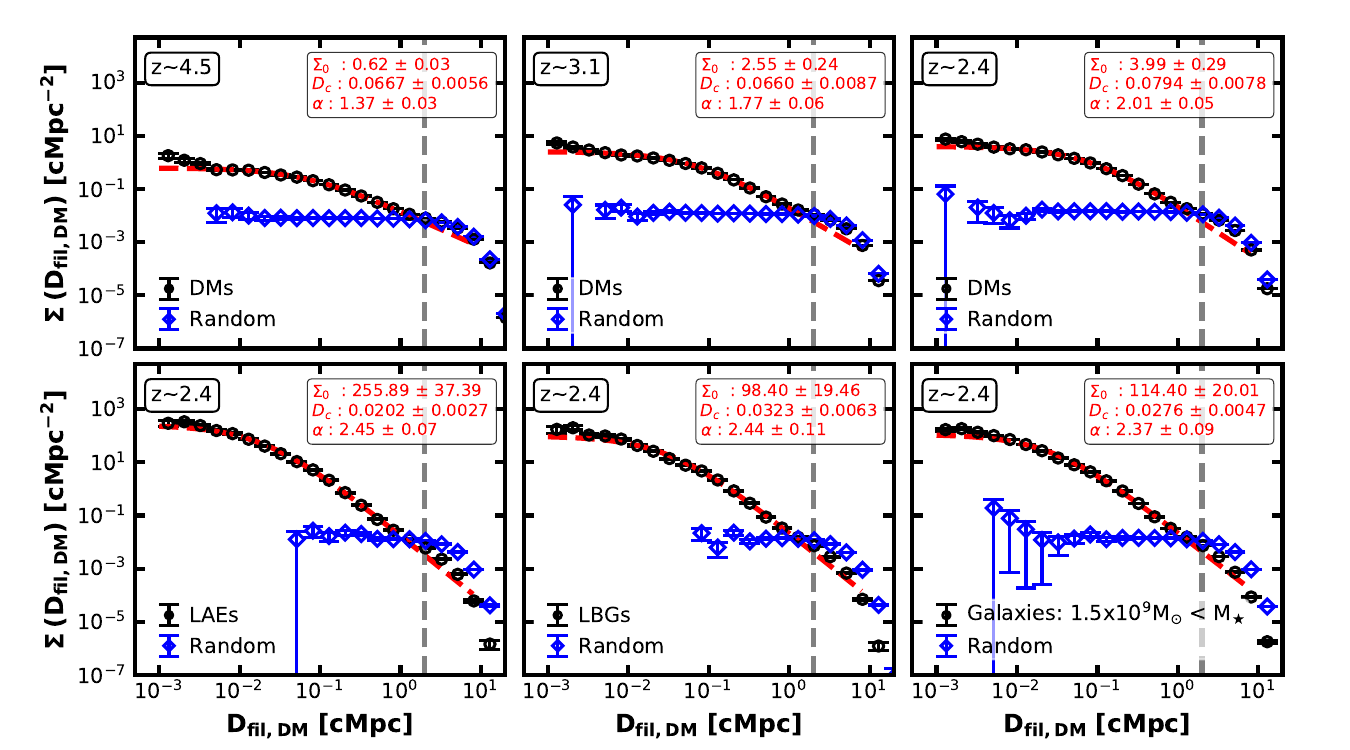}
    \caption{Fitting results for the vertical density profile of dark matter particles at three redshifts (upper panels) and LAEs, LBGs, and galaxies with stellar masses larger than $1.5 \times {10}^9 \;\rm{M_{\odot}}$ at $z\sim2.4$ (lower panels). We also present the best-fit values of ${\Sigma}_0$, $D_c$, and $\alpha$ for each sample at the upper right corner of each panel. The gray dashed line in each panel represents the maximum fitting range ($2\;\rm{cMpc}$). The blue markers show the vertical density profile of randomly distributed particles with the same size as the sample at each panel. The empirical fitting formula of equation (\ref{eq:empirical_formula}) shows good agreement at $D_{fil,\rm{DM}} \lesssim 2 \; \rm{cMpc}$.}
    \label{fig:representative_fitting_results}
\end{figure*}

To examine how the choice of the minimum persistence level affects our results, we also use a higher value ($10\sigma$) and perform similar analyses in Appendix \ref{appendix:10sigma_results}. We also use another filament finder T-ReX \citep{Bonnaire_2020} to examine whether there is any bias introduced by using only one filament finder and show the results in Appendix \ref{appendix:from_TReX}. In both cases, the median values become higher than those in Figure \ref{fig:median_Dskel_comparison}. However, both tests show trends similar to Figure \ref{fig:median_Dskel_comparison}, indicating the robustness of our results to the choice of the persistence level and the filament finder.

\subsection{Empirical Fitting Formula for Vertical Density Profiles}\label{subsec:fitting_formula}

To have a quantitative description of the vertical density profile around dark matter filaments, $\Sigma \, (D_{fil,\rm{DM}})$, we introduce the following fitting formula:
\begin{equation}\label{eq:empirical_formula}
\Sigma \, (D_{fil,\rm{DM}}) = \frac{{\Sigma}_0}{{(1 + D_{fil,\rm{DM}}/D_c)}^\alpha}.
\end{equation}
Here, ${\Sigma}_0$ is the density at the center of the filamentary structure (i.e., where $D_{fil,\rm{DM}} = 0$) and $D_c$ is a characteristic distance from the filament where the slope of the profile changes from zero to $-\alpha$. 

The density profiles for our samples of LAEs, LBGs, all galaxies, and dark matter particles are well fitted with the equation (\ref{eq:empirical_formula}) within $D_{fil,\rm{DM}} \lesssim 2 \; \rm{cMpc}$ regardless of redshift. We present the example cases with the best-fit curves in Figure \ref{fig:representative_fitting_results}. Here, we show only the results for dark matter particles at three redshifts (upper panels) and LAEs, LBGs, and galaxies with stellar masses larger than $1.5 \times 10^9 \; \rm{M_{\odot}}$ at $z\sim2.4$ (lower panels). We also generate comparison samples of randomly distributed particles with the same size as each sample and calculate their $\Sigma\,(D_{fil,\rm{DM}})$. The density profile of these randomly distributed samples starts to decrease at some point because of the mean separation between the filaments. When $D_{fil,\rm{DM}}\gtrsim 2 \; \rm{cMpc}$, the density profiles for all of our samples become smaller than those of randomly distributed particles (see Figure \ref{fig:representative_fitting_results}). This can be interpreted as the size of the overdense region around dark matter filaments in our study is $\sim2\;\rm{cMpc}$. This size of the overdense region could vary with the choice of the minimum persistence level. The fitting results for all of our samples are shown in Appendix \ref{appendix:fitting_results}.

\subsection{Spatial Distributions of Samples Around Filaments Defined by Themselves}\label{subsec:thickness}

We also examine the spatial distributions of DMs, LAEs, LBGs, and all galaxies around the filamentary structures defined by themselves. The shape of the extracted filamentary structures depends not only on the choice of the minimum persistence level, but also on the size of the sample. Therefore, we generate 1,000 subsamples with the same size of 7,000 for each sample to make a fair comparison between them. The size of the subsamples is chosen to be slightly smaller than the size of the LAE samples at $z\sim4.5$. Each subsample is constructed from a random selection of the members from each galaxy sample. We extract the filamentary structures of these random subsamples with a minimum persistence level of $3\sigma$. We again calculate the distance from a galaxy or a dark matter particle in each sample to the closest filament defined by themselves; $D_{fil,\rm{self}}$. Here, the galaxies and dark matter particles close to the nodes ($\leq 2 \; \rm{cMpc}$) are ignored to remove the effect of the nodes. We calculate median values of $D_{fil,\rm{self}}$ for 1,000 random subsamples of each sample. Then, we calculate the mean and the standard deviation value of those 1,000 median $D_{fil,\rm{self}}$ values for each sample.

\begin{figure*}[t!]
    \centering
    \includegraphics[width=\textwidth]{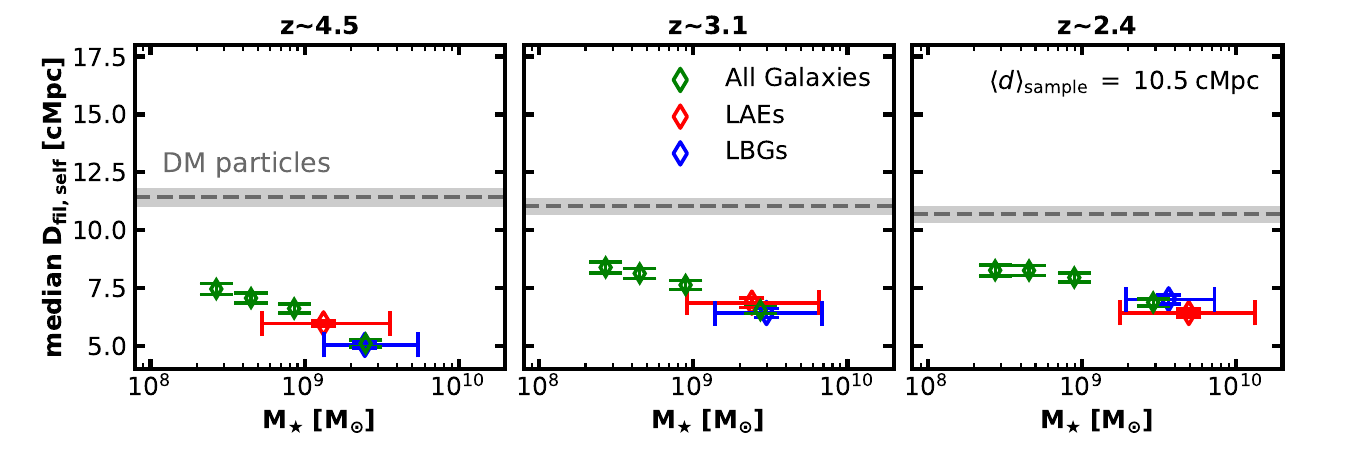}
    \caption{Median $D_{fil,\rm{self}}$ value of each sample at three redshifts. The shaded regions and vertical errorbars show the standard deviations for the samples (see text for detail), while the horizontal errorbars for LAEs and LBGs represent 15.9 and 84.1 percentile of stellar masses. The galaxy samples including LAEs and LBGs show smaller median values than dark matter particles. We also present the mean separation of the sample used for extracting filaments at the right panel.}
    \label{fig:thickness_of_filaments}
\end{figure*}

Figure \ref{fig:thickness_of_filaments} shows the results for the LAEs, LBGs, all galaxies, and dark matter particles at three redshifts. Here, we present the standard deviation for each sample as a shaded region or a vertical errorbar. At all redshifts, LAEs, LBGs, and all galaxies have smaller median $D_{fil,\rm{self}}$ values than dark matter particles. This result implies that LAEs, LBGs, and all galaxies form thinner filamentary structures than dark matter. We also find that more massive galaxies tend to be more concentrated toward their filaments than less massive galaxies, which is consistent with the result of Section \ref{subsec:Dskel_dist_DM_filament}. LAEs and LBGs again follow this stellar-mass trend well. Similar to Figure \ref{fig:median_Dskel_comparison}, the absolute values of the median $D_{fil,\rm{self}}$ of the samples would change if one uses a different minimum persistence level or different sample size. Therefore, we present the mean separation of the samples used to extract the filaments (here, each subsample with a size of 7,000) in the right panel of Figure \ref{fig:thickness_of_filaments}.

\begin{figure*}[t!]
    \centering
    \includegraphics[width=\linewidth]{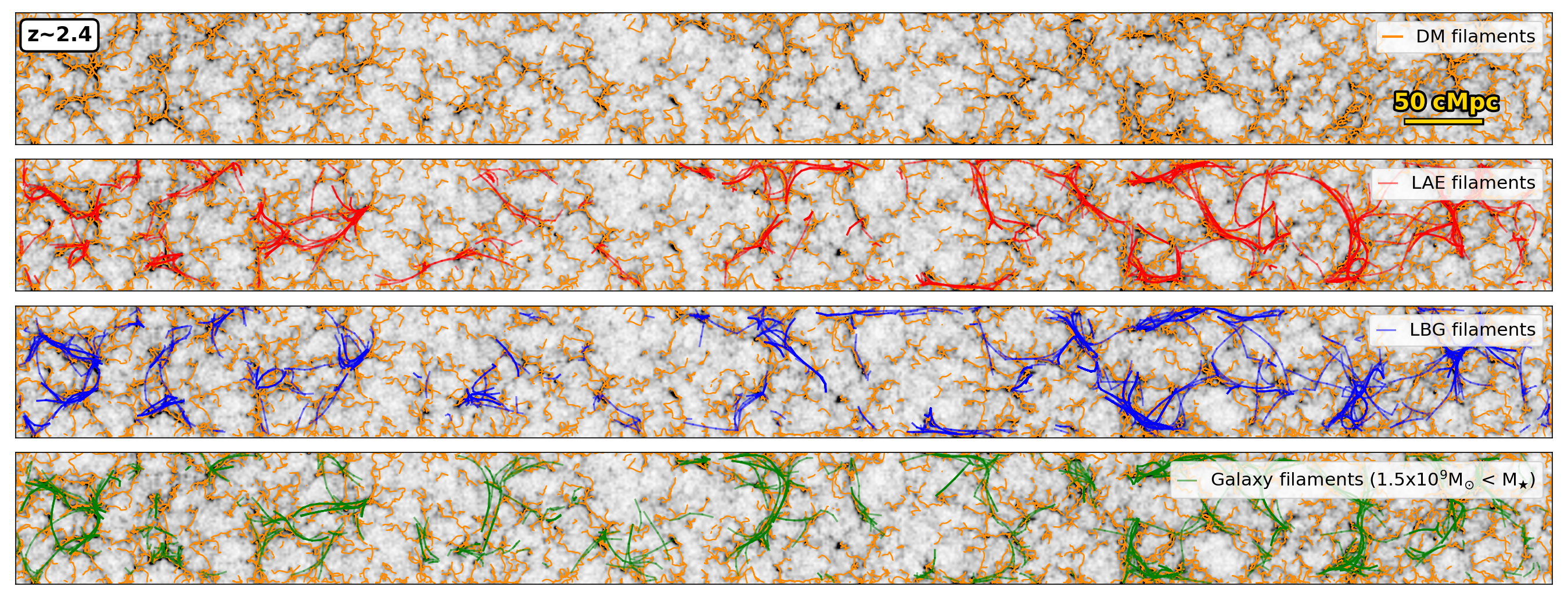}
    \caption{Two-dimensional view of filamentary structures from 5 random subsamples of LAEs, LBGs, and all galaxies with stellar masses larger than $1.5 \times {10}^9 \; \rm{M_{\odot}}$ at $z\sim2.4$. We present the distribution of dark matter and their filamentary structures as background in each panel. For comparison, we plot the filaments within the same region of Figure \ref{fig:2d_distribution_of_samples}.}
    \label{fig:2D_filaments_ndata7000}
\end{figure*}

\begin{figure*}
    \centering
    \includegraphics[width=\textwidth]{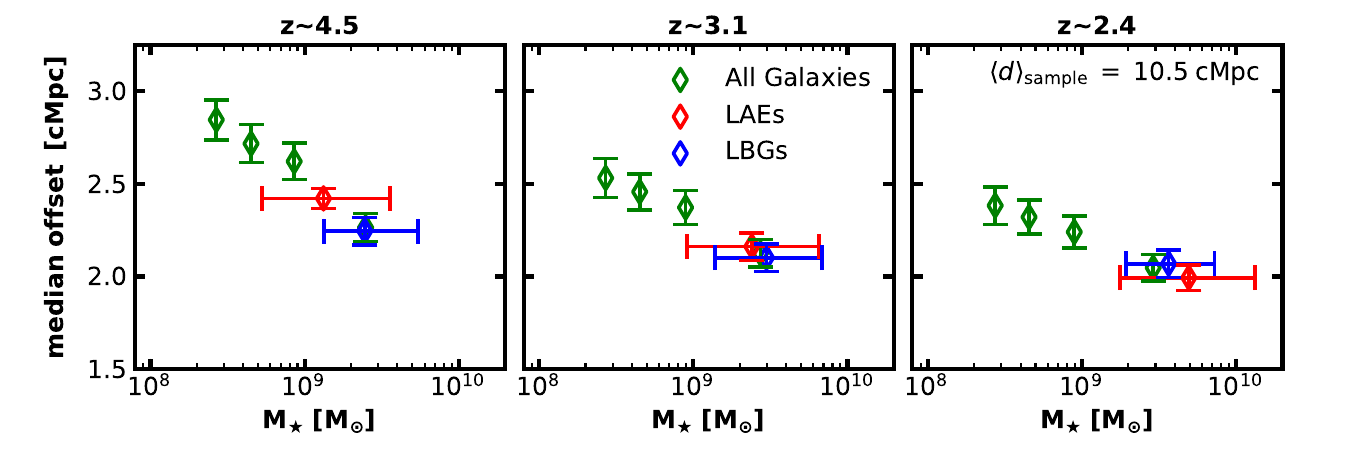}
    \caption{Median spatial offset between filamentary structures traced by galaxy samples (i.e., LAEs, LBGs, and all galaxies) and dark matter particles. Here we also present the mean separation of the sample, which is used for extracting filamentary structures. The vertical errorbars represent the standard deviation of mean offsets from 1,000 different random subsamples, while the horizontal errorbars for LAEs and LBGs represent their 15.9 and 84.1 percentile of stellar mass.}
    \label{fig:mean_separation}
\end{figure*}

\subsection{Difference between Filamentary Structures from Dark Matter Particles and Galaxies}\label{subsec:difference_between_filaments}

We examine whether there is any systematic spatial offset between the filamentary structures defined by dark matter particles and those by LAEs, LBGs, and all galaxies. We again use the filamentary structures from the 1,000 randomly selected subsamples of Section \ref{subsec:thickness} for LAEs, LBGs, and all galaxies. We then compare them to the dark matter filaments in Section \ref{subsec:Dskel_dist_DM_filament}. Figure \ref{fig:2D_filaments_ndata7000} shows a two-dimensional projected view of these filamentary structures at the same region of Figure \ref{fig:2d_distribution_of_samples} at $z\sim2.4$. For convenience, we present only 5 cases out of 1,000 subsamples constructed from each galaxy sample with random selection. The filamentary structures of the galaxy samples show good agreement with those of dark matter particles in overall shapes. However, there are discrepancies in details between filamentary structures traced by dark matter particles and those by LAEs, LBGs, and all galaxies. Especially, there are many missing dark matter filaments when we identify filaments using galaxy samples. It would be because of the nature of the galaxies, which are biased tracers of dark matter distributions, and the effect of the random realization of the galaxy samples.

To quantify the difference between the filamentary structures of the galaxy samples and the dark matter particles, we calculate the median spatial offset between them. As mentioned in Section \ref{subsec:extracting_filaments}, the filamentary structures extracted by DisPerSE consist of small segments. We calculate the distance from each segment of the filamentary structures defined by galaxy samples (i.e., LAEs, LBGs, and all galaxies) to the filamentary structures defined by dark matter particles. We calculate the median value of these distances to refer to the median spatial offset between the filamentary structures of galaxy samples and dark matter. 

Figure \ref{fig:mean_separation} shows the median offsets for three redshifts. Here, the vertical errorbars represent the standard deviation of the median offset values from 1,000 random subsamples of our samples. As in Figures \ref{fig:median_Dskel_comparison} and \ref{fig:thickness_of_filaments}, we present the mean separation of the sample, which is used for extracting filamentary structures. Galaxy samples, including LAEs and LBGs, have median values smaller than the mean separation of them (i.e., $10.5 \; \rm{cMpc}$). As in Sections \ref{subsec:Dskel_dist_DM_filament} and \ref{subsec:thickness}, there is a trend that massive galaxies have smaller median offset values than less massive galaxies. This is because more massive galaxies reside in denser regions than less massive galaxies. LAEs and LBGs again follow the stellar-mass trend well, implying that they have no systematic bias compared to the entire galaxy population. We also perform a similar analysis for random subsamples with a larger sample size and find a similar result (see Appendix \ref{appendix:different_ndata}).

\section{Discussion}\label{sec:discussion}

We perform three sets of analyses of the filamentary structures traced by LAEs, LBGs, all galaxies, and dark matter particles in Section \ref{sec:results}. Here, we discuss the results of those analyses. First, in Section \ref{subsec:comparison_with_observations}, we compare our results with previous studies based on observations and simulations. In Section \ref{subsec:capability_of_LAEs_LBGs}, we discuss the capability of LAEs and LBGs as tracers of large-scale structures of dark matter at high redshifts.

\subsection{Comparison with other Studies}\label{subsec:comparison_with_observations}

We first focus on the relation between the stellar mass of galaxies and their distance from filaments. Many observational studies examine how filamentary structures affect the growth of stellar mass in galaxies. For instance, \cite{Kraljic2017} analyze the local $(0.03 \leq z \leq 0.25)$ galaxies from the Galaxy And Mass Assembly \citep[GAMA:][]{Driver_2009} redshift survey to examine how galaxy properties change within filamentary structures. They find that the galaxies with larger stellar masses are more concentrated toward the filaments than those with smaller stellar masses. \cite{Laigle_2018} also find similar results with galaxies from the COSMOS2015 \citep{Laigle_2018} photometric redshift catalog at a redshift range $0.5 < z < 0.9$. Some studies based on simulations also find a similar relation between the mass of halos and their distance to filaments (e.g., \citealt{Jhee_2022}). Our simulation-based results in Sections \ref{subsec:Dskel_dist_DM_filament} and \ref{subsec:thickness} are consistent with these results. These results also support the idea of the hierarchical structure in the sense that more massive galaxies would be formed and evolved in denser regions of dark matter. 

On the other hand, there are some observational studies focusing on the relationship between the star formation activity of galaxies and their distance from filaments. Again, \cite{Kraljic2017} and \cite{Laigle_2018} find that passive galaxies reside closer to the filamentary structures than star-forming galaxies in the local universe. This environment dependence is expected to be reversed at high redshifts because of larger amount of gas in high-redshift galaxies (i.e., the reversal of the star formation rate-density relation, \citealt{Elbaz_2007}; \citealt{Hwang_2019}; \citealt{Song_2021}; \citealt{Martin-Navarro_2023_arxiv}). Indeed, we could see a hint of such a trend (not shown here), which needs more detailed analysis by simultaneously considering the mass effect as well; this will be the topic of future studies.

\subsection{LAEs and LBGs as Tracers of Large-Scale Structures}\label{subsec:capability_of_LAEs_LBGs}

We discuss the capability of LAEs and LBGs as tracers of large-scale structures of dark matter at high redshifts. In Section \ref{subsec:Dskel_dist_DM_filament}, we examine the spatial distribution of our samples around the filamentary structures of dark matter particles and find that LAEs and LBGs are more concentrated toward the dark matter filaments than dark matter particles themselves. We also examine the spatial distribution of our samples around the filamentary structures defined by themselves in Section \ref{subsec:thickness}. We find that LAEs and LBGs are again more concentrated toward filaments than dark matter. In both Sections \ref{subsec:Dskel_dist_DM_filament} and \ref{subsec:thickness}, we find a stellar-mass trend that massive galaxies are more concentrated toward filaments. This trend could be understood by the fact that massive galaxies have larger galaxy bias (i.e., higher probability to be formed in dense regions) than less massive ones \citep[e.g.,][]{Kaiser_1984, Mo_White_1996}. It is important to note that LAEs and LBGs follow this stellar-mass trend of the entire galaxy population well without systematic bias. These results, combined with the fact that they are preferred tracers in observations, suggest that LAEs and LBGs can be good tracers of filamentary structures of dark matter at high redshifts.

In Section \ref{subsec:difference_between_filaments}, we also analyze the difference between filamentary structures traced by galaxy samples (i.e., LAEs, LBGs, and all galaxies with various stellar mass ranges) and dark matter particles. The filamentary structures traced by galaxy samples look similar to those by dark matter particles in terms of overall shape. We also calculate the spatial offsets between filamentary structures and find that the median offsets of the galaxy samples are smaller than the mean separation of the sample. We also find a trend that massive galaxies have smaller median offsets than less massive galaxies. LAEs and LBGs again seem to follow this mass trend, suggesting that there is no systematic bias of LAEs and LBGs as tracers of filamentary structures of dark matter at high redshifts compared to the entire galaxy population. However, it is important to note that galaxy samples including LAEs and LBGs cannot trace all the features of large-scale structures of dark matter. We confirm that galaxies including LAEs and LBGs can trace the overall shape of dark matter filaments with spatial offsets smaller than the mean separation of them.

Furthermore, the LBG samples from each snapshot data in this study cannot be directly compared to the photometrically selected LBGs in observations, which have wide redshift ranges ($\Delta z \sim0.5$). These wide redshift ranges would make it difficult to trace three-dimensional large-scale structures using the projected distribution of LBGs from observations. On the other hand, LAEs from narrow-band observations have narrower redshift ranges ($\Delta z \lesssim 0.04-0.08$). Therefore, it would be more efficient to use LAEs to study high-z large-scale structures in observations, such as the ODIN survey, than to use LBGs.

\section{Conclusions}\label{sec:conclusion}

We have tested the capability of LAEs and LBGs as tracers of large-scale structures of dark matter at high redshifts using the data from the Horizon Run 5 cosmological hydrodynamical simulation at three redshifts: $z\sim2.4$, $3.1$, and $4.5$. To do that, we have generated samples of LAEs, LBGs, all galaxies in various stellar mass ranges, and dark matter particles. We have focused on the filamentary structures of these samples, which were extracted by the DisPerSE code. We have performed three different tests on the spatial distributions and filamentary structures of these samples and found the following results.

\begin{enumerate}

    \item [(I)] We have examined the spatial distributions of our samples of LAEs, LBGs, all galaxies, and dark matter particles around the filamentary structure defined by dark matter particles. The result shows that LAEs and LBGs are more concentrated toward the dark matter filaments than dark matter particles. We also find that more massive galaxies are more concentrated than less massive galaxies, which is consistent with the previous results from observations and simulations. 
    
    \item [(II)] We find an empirical fitting formula as equation (\ref{eq:empirical_formula}) for the vertical density profile around the filamentary structures. The formula works well within $2\;\rm{cMpc}$ from the filamentary structures. Beyond $2\;\rm{cMpc}$, the density profiles for all of our samples become smaller than that of randomly distributed particles.
    
    \item [(III)] We have compared the spatial distribution of our samples around the filamentary structure defined by themselves. Again, LAEs and LBGs are more concentrated toward their filaments than dark matter, suggesting that they form thinner filamentary structures than dark matter. Furthermore, more massive galaxies are again more concentrated toward their filaments. 
    
    \item [(IV)] The filamentary structures traced by galaxy samples (i.e., LAEs, LBGs, and all galaxies) and dark matter particles show good agreement in terms of their overall shapes. The median spatial offsets between the filamentary structures of galaxy samples and dark matter particles are smaller than the mean separation of the galaxy sample used for extracting filaments. We also find a stellar-mass trend that more massive galaxies have smaller median spatial offsets from dark matter filaments.
    
    \item [(V)] We also find that our LAE and LBG samples follow well the stellar-mass trend in the results (I), (II), and (IV). Although the stellar-mass seems to be a dominant factor for the capability as the tracers of dark matter filaments, our results suggest that there is no systematic bias of LAEs and LBGs as such tracers compared to the entire galaxy population. 
    
\end{enumerate}

These results support the idea that LAEs and LBGs could be good tracers of large-scale structures of dark matter at high redshifts. It should be noted that we have analyzed three-dimensional distributions of LAEs and LBGs in Horizon Run 5. Because photometric surveys like the ODIN survey will provide the data for the two-dimensional distributions of LAEs and LBGs, we need to analyze the two-dimensional mock data of HR5 and compare it with the observational results. Furthermore, to generate more realistic LBG samples, we need the lightcone data of cosmological simulations with continuous redshift range. We plan to construct the lightcone data of HR5 and will perform the analyses similar to this study and compare the results with the ODIN observed data.

\section*{Acknowledgments}
We thank the referee for constructive comments that improved the manuscript. HSH acknowledges the support by the National Research Foundation of Korea (NRF) grant funded by the Korea government (MSIT), NRF-2021R1A2C1094577 and by Samsung Electronic Co., Ltd. (Project Number IO220811-01945-01).
We thank the support from the KISTI National Supercomputing Center and its Nurion Supercomputer through the Grand Challenge Program (KSC-2018-CHA-0003, KSC-2019-CHA-0002).
Large data transfer was supported by KREONET, which is managed and operated by KISTI. This work is also supported by the Center for Advanced Computation at Korea Institute for Advanced Study. JP was supported by a KIAS Individual Grant (PG078702) at Korea Institute for Advanced Study. J.L. is supported by the National Research Foundation of Korea (NRF-2021R1C1C2011626). This work was also partially supported by the National Research Foundation of Korea (NRF) grant funded by the Korea government (MSIT, 2022M3K3A1093827). This work was partially supported by the National Research Foundation of Korea (NRF) grant funded by the Korea government (MSIT) (2022M3K3A1093827). 
YK is supported by Korea Institute of Science and Technology Information (KISTI) under the institutional R\&D project (K24L2M1C4).
EG would also like to acknowledge support from NSF grant AST-2206222.
ONS acknowledges support from ERC Consolidator Grant ``GAIA-BIFROST" (Grant Agreement ID 101003096) and the UK Science and Technology Facilities Council (Consolidated Grant ST/V000721/1; Small Award ST/Y002695/1).

\appendix

\begin{figure*}
    \centering
    \includegraphics[width=\textwidth]{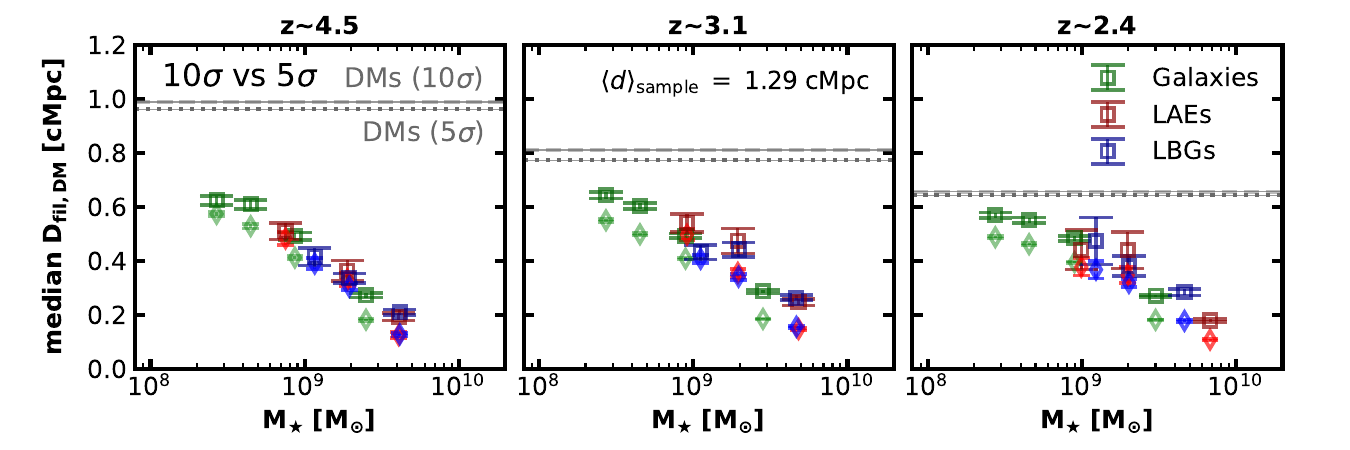}
    \caption{Median $D_{fil,\rm{DM}}$ value of each sample at three redshifts with $10\sigma$ (darker square) and $5\sigma$ (fainter diamond markers) minimum persistence levels. The meaning of the errorbars and shaded regions are the same as in Figure \ref{fig:median_Dskel_comparison}.}
    \label{fig:median_Dskel_comparison_nsig10}
\end{figure*}

\begin{figure*}
    \centering
    \includegraphics[width=\textwidth]{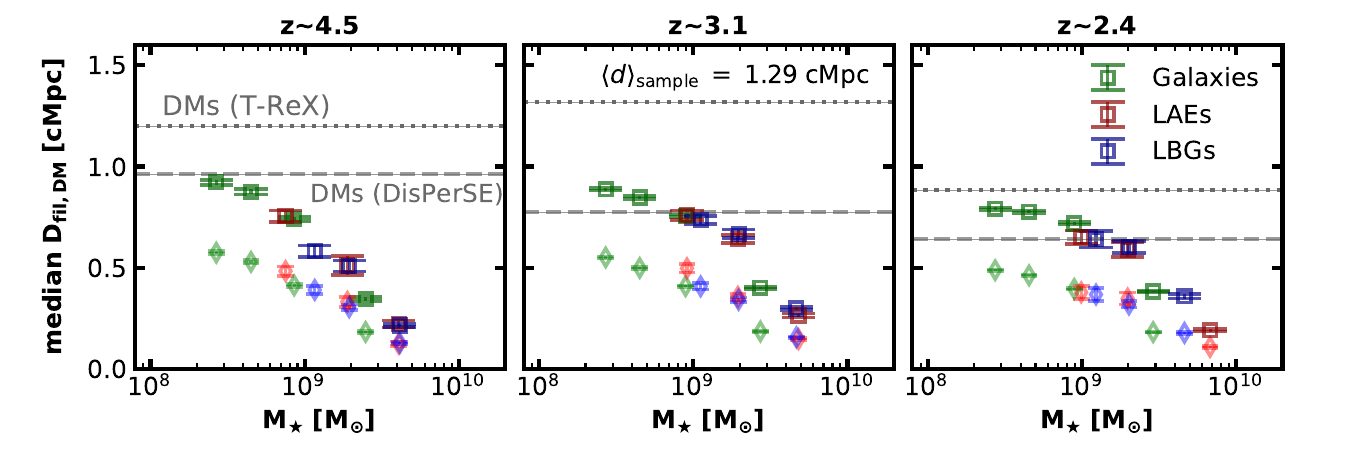}
    \caption{Median $D_{fil,\rm{DM}}$ value of each sample at three redshifts using T-Rex (darker square) and DisPerSE (fainter diamond markers) as the filament finder. The meaning of the errorbars and shaded regions are the same as in Figure \ref{fig:median_Dskel_comparison}.}
    \label{fig:median_dskel_TReX}
\end{figure*}

\section{Results from a higher minimum persistence level}\label{appendix:10sigma_results}

We adopt $5\sigma$ as the minimum persistence level when we extract filamentary structures of dark matter particles in Section \ref{subsec:extracting_filaments}. Here, we present how the choice of the minimum persistence level affects the results of our analysis in Sections \ref{sec:results}. We calculate again the median values of $D_{fil,\rm{DM}}$ of our samples with the filamentary structures of dark matter extracted with $10\sigma$ minimum persistence level. As mentioned in Section \ref{subsec:extracting_filaments}, filaments with higher persistence levels represent more robust and less detailed structures. We present the results in Figure \ref{fig:median_Dskel_comparison_nsig10} along with those of $5\sigma$ level. The absolute values of median $D_{fil,\rm{DM}}$ become larger ($\Delta D_{fil,\rm{DM}} \lesssim 0.1 \; \rm{cMpc}$) than those with $5\sigma$ level, because the mean separation between filaments becomes larger. Again, LAEs, LBGs, and all galaxies have smaller median values than dark matter particles. We also find LAEs and LBGs follow well the stellar-mass trend of all galaxies as in Section \ref{subsec:Dskel_dist_DM_filament}. These indicate that the choice of the minimum persistent level of filamentary structures would not affect our main results.

\section{Results with T-ReX filament finder}\label{appendix:from_TReX}

We also perform a similar analysis to Section \ref{subsec:Dskel_dist_DM_filament} with the filamentary structures extracted with T-ReX filament finder \citep{Bonnaire_2020}. T-ReX finds filamentary structures as the ridges of the given data points (here, galaxies or dark matter particles) using graph theory and the Gaussian Mixture Model. We would like to emphasize that the mathematical definition of filaments in T-ReX is different from that in DisPerSE. Therefore, comparing the results of the two algorithms could show the robustness of our results with respect to how the filaments are defined and identified. A comprehensive comparison between various filament finders is available in \cite{Libeskind_2018}. 

Figure \ref{fig:median_dskel_TReX} compares the median value of $D_{fil,\rm{DM}}$ of the samples using the dark matter filaments from T-ReX and DisPerSE. Although the absolute values of the median $D_{fil,\rm{DM}}$ are larger than those with DisPerSE filaments ($\Delta D_{fil,\rm{DM}} \lesssim 0.5 \; \rm{cMpc}$), the overall trend is similar. This result shows that our main conclusions would not change much when we use different filament finders for our analyses.

\section{Fitting results for all samples}\label{appendix:fitting_results}

\begin{figure*}[t!]
    \centering
    \includegraphics[width=\textwidth]{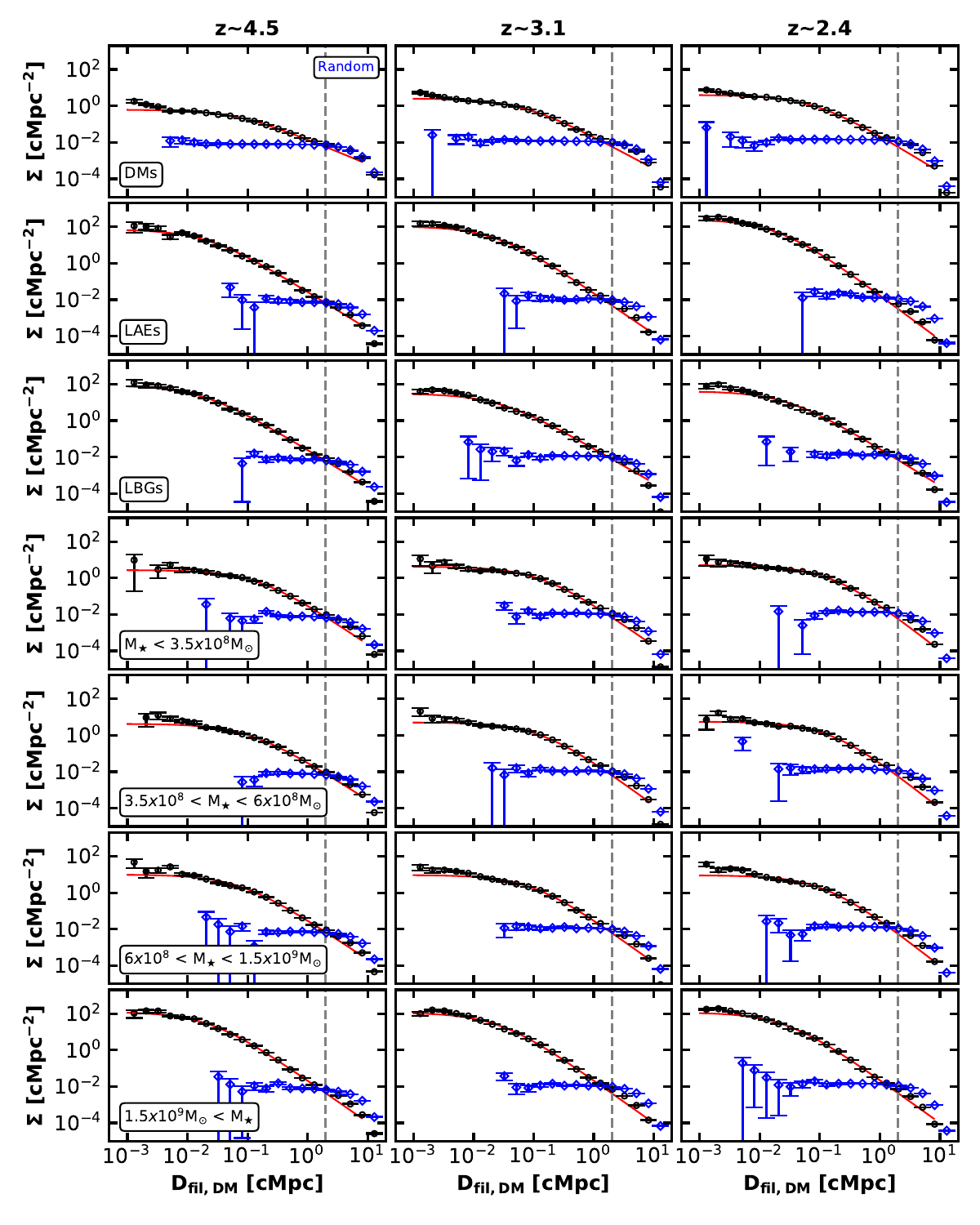}
    \caption{The vertical density profiles for all of our samples (black markers) and random particles (blue markers) for the case of the $5\sigma$ minimum persistence level. Above $\sim2\;\rm{cMpc}$ (gray dashed line in each panel), the density profiles of our samples become smaller than those of random particles. The red solid line in each panel represents the best-fit curve for each sample using the equation (\ref{eq:empirical_formula}).}
    \label{fig:fitting_results_nsig5}
\end{figure*}

In Section \ref{subsec:fitting_formula}, we find an empirical fitting formula for the vertical density profile around filamentary structures as equation \ref{eq:empirical_formula}. Here, we present the fitting results of all of our samples at three redshifts in Figure \ref{fig:fitting_results_nsig5}. The results show that our empirical formula works well for all of our samples at three redshifts within $\sim 2\; \rm{cMpc}$ regardless of redshift. We also present the best-fit values for the fitting parameters (i.e., $\Sigma_{0}$, $D_{c}$, and $\alpha$) in Table \ref{table:bestfit_values}.

\begin{figure*}[t!]
    \centering
    \includegraphics[width=\textwidth]{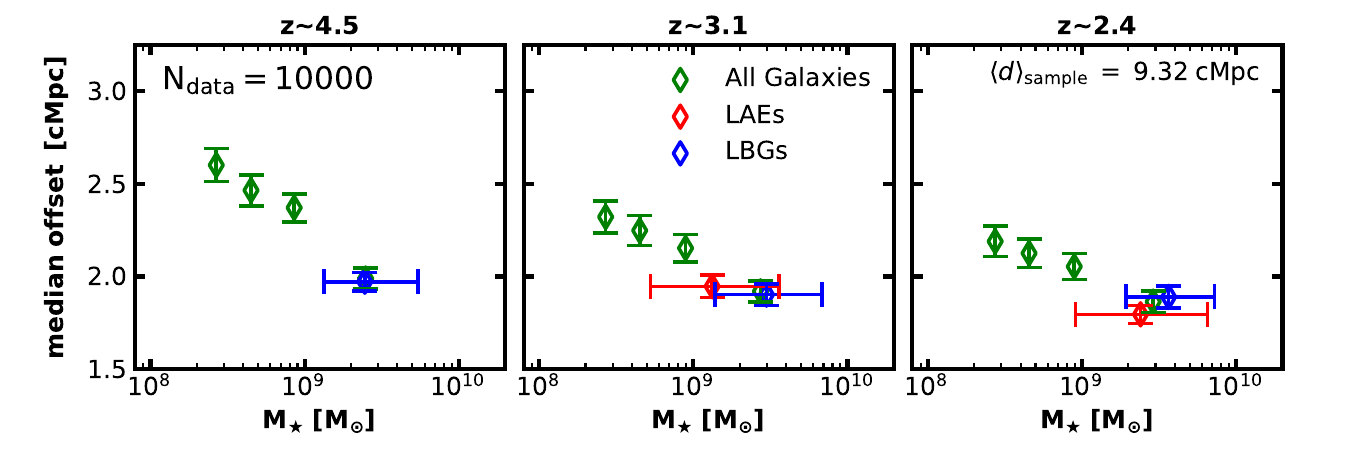}
    \caption{Median spatial offset between filamentary structures traced by galaxy samples (i.e., LBGs and all galaxies of several stellar mass ranges) and dark matter particles. The errorbars and shaded regions represent the standard deviation of mean offsets from 1,000 different random subsamples. The mean offset values of the samples are similar to each other and smaller than the mean separation of them.}
    \label{fig:mean_separation_10000}
\end{figure*}

\begin{deluxetable*}{c|c|lll|llll}
    \tabletypesize{\scriptsize}
    \renewcommand{\arraystretch}{1.5}
    \renewcommand{\tabcolsep}{1.5mm}
    \tablecaption{Best-fit values of the parameters in the fitting formula for the vertical density distributions in Section \ref{sec:results}.}
    \label{table:bestfit_values}
    \tablehead{\colhead{} & \colhead{} & \colhead{} & \colhead{} & \colhead{} & \multicolumn{4}{c}{All galaxies} \\ \colhead{redshift} & \colhead{parameters} & \colhead{DM particles} & \colhead{LAEs} & \colhead{LBGs} & \colhead{bin 1} & \colhead{bin 2} & \colhead{bin 3} & \colhead{bin 4}}
    \startdata
    \hline
    \multirow{3}{*}{$z\sim2.4$} & $\Sigma_{0}\;[\rm{cMpc^{-2}}]$ & 3.99$\;\pm\;$0.29 & 256$\;\pm\;$37 & 98.4$\;\pm\;$19.5 & 5.02$\;\pm\;$0.48 & 5.60$\;\pm\;$0.57 & 8.98$\;\pm\;$1.18 & 114$\;\pm\;$20 \\
    & $D_{c}\;\rm{[cMpc]}$ & 0.079$\;\pm\;$0.008 & 0.020$\;\pm\;$0.003 & 0.032$\;\pm\;$0.006 & 0.158$\;\pm\;$0.024 & 0.155$\;\pm\;$0.025 & 0.120$\;\pm\;$0.022 & 0.028$\;\pm\;$0.005 \\
    & $\alpha$ & 2.01$\;\pm\;$0.05 & 2.45$\;\pm\;$0.07 & 2.44$\;\pm\;$0.12 & 2.59$\;\pm\;$0.15 & 2.62$\;\pm\;$0.16 & 2.58$\;\pm\;$0.16 & 2.37$\;\pm\;$0.09 \\
    \hline
    \multirow{3}{*}{$z\sim3.1$} & $\Sigma_{0}\;[\rm{cMpc^{-2}}]$ & 2.55$\;\pm\;$0.24 & 106$\;\pm\;$16 & 114$\;\pm\;$18 & 4.17$\;\pm\;$0.42 & 5.13$\;\pm\;$0.46 & 9.31$\;\pm\;$1.03 & 111$\;\pm\;$18 \\
    & $D_{c}\;\rm{[cMpc]}$ & 0.066$\;\pm\;$0.009 & 0.024$\;\pm\;$0.003 & 0.026$\;\pm\;$0.004 & 0.133$\;\pm\;$0.021 & 0.134$\;\pm\;$0.018 & 0.098$\;\pm\;$0.015 & 0.027$\;\pm\;$0.004 \\
    & $\alpha$ & 1.77$\;\pm\;$0.06 & 2.26$\;\pm\;$0.06 & 2.32$\;\pm\;$0.08 & 2.31$\;\pm\;$0.12 & 2.42$\;\pm\;$0.11 & 2.39$\;\pm\;$0.11 & 2.36$\;\pm\;$0.08 \\
    \hline
    \multirow{3}{*}{$z\sim4.5$} & $\Sigma_{0}\;[\rm{cMpc^{-2}}]$ & 0.62$\;\pm\;$0.03 & 71.1$\;\pm\;$18.4 & 125$\;\pm\;$20 & 2.71$\;\pm\;$0.18 & 4.27$\;\pm\;$0.61 & 9.82$\;\pm\;$1.79 & 122$\;\pm\;$20 \\
    & $D_{c}\;\rm{[cMpc]}$ & 0.067$\;\pm\;$0.006 & 0.022$\;\pm\;$0.005 & 0.023$\;\pm\;$0.004 & 0.149$\;\pm\;$0.017 & 0.106$\;\pm\;$0.022 & 0.075$\;\pm\;$0.017 & 0.024$\;\pm\;$0.004 \\
    & $\alpha$ & 1.37$\;\pm\;$0.03 & 2.04$\;\pm\;$0.10 & 2.30$\;\pm\;$0.07 & 2.22$\;\pm\;$0.09 & 2.11$\;\pm\;$0.14 & 2.19$\;\pm\;$0.14 & 2.30$\;\pm\;$0.07 \\
    \hline
    \enddata
\end{deluxetable*}


\section{Results Random Subsamples with Larger Sample Size}\label{appendix:different_ndata}

In Section \ref{subsec:difference_between_filaments}, we analyze the median spatial offset values between filamentary structures traced by galaxy samples and dark matter particles using randomly selected subsamples with a size of 7,000. Here, we perform a similar analysis but for the subsamples constructed with a larger sample size of 10,000. We exclude the LAE sample at $z\sim4.5$ which has a smaller sample size than 10,000. Figure \ref{fig:mean_separation_10000} shows the median spatial offsets between filamentary structures of galaxy samples and dark matter. Again, we find that the median offsets are smaller than the mean separation of the sample and the stellar-mass trend that massive galaxies show smaller offsets than less massive ones. LAEs and LBGs follow well this trend, indicating that they have no systematic bias as tracers of filamentary structures of dark matter compared to the entire galaxy population.

\bibliography{refs}{}
\bibliographystyle{aasjournal}

\end{document}